\documentclass[%
 reprint,
 amsmath,amssymb,
 aps, superscriptaddress
]{revtex4-2}

\usepackage{graphicx}
\usepackage{dcolumn}
\usepackage{bm}
\usepackage{amsmath}
\usepackage{physics}
\usepackage{slashed}
\usepackage{float}
\usepackage{dsfont}
\usepackage{gensymb}
\usepackage{comment}
\usepackage[thinlines]{easytable}
\usepackage[dvipsnames]{xcolor}
\usepackage{multirow}
\usepackage{array}    
\usepackage{booktabs} 
\usepackage{lipsum}
\usepackage{bbold}
\usepackage{mathtools}
\usepackage{csquotes}

\usepackage{hyperref}
\hypersetup{
    colorlinks=true, 
    linktoc=all, 
    linkcolor=blue, 
    citecolor=blue}

\usepackage{etoolbox}

\AtBeginEnvironment{tabular}{}
\AtBeginEnvironment{easytable}{}

\allowdisplaybreaks

\newcommand{\sbra}[2]{%
  \prescript{}{#1}{\bra{#2}}%
}

\newcommand{\sket}[2]{%
  \ket{#2}_{#1}%
}

\newcommand{\sbraket}[4]{%
  \prescript{}{#1}{\braket{#2}{#3}}_{#4}%
}

\begin{document}


\title{Optimizing the description of the Delta region in the Ghent Hybrid model for single-pion production}

\author{M.~Hooft}
\affiliation{Department of Physics and Astronomy, Ghent University, B-9000 Gent, Belgium}

\author{A.~Nikolakopoulos}
\affiliation{Department of Physics and Astronomy, Ghent University, B-9000 Gent, Belgium}
\affiliation{Department of Physics, University of Washington, Seattle, WA 98195-1560, USA}

\author{J.~García-Marcos}
\affiliation{Department of Physics and Astronomy, Ghent University, B-9000 Gent, Belgium}
\affiliation{Grupo de Física Nuclear,
Departamento de Estructura de la Materia, Física Térmica y Electrónica,
Facultad de Ciencias Físicas, Universidad Complutense de Madrid and IPARCOS
CEI Moncloa, Madrid 28040, Spain}

\author{Y.~De~Backer}
\affiliation{Department of Physics and Astronomy, Ghent University, B-9000 Gent, Belgium}

\author{T.~Franco-Munoz}
\affiliation{Department of Physics and Astronomy, Ghent University, B-9000 Gent, Belgium}

\author{K.~Niewczas}
\affiliation{Department of Physics and Astronomy, Ghent University, B-9000 Gent, Belgium}

\author{R.~González-Jiménez}
\affiliation{Departamento de Física Atómica, Molecular y Nuclear, Universidad de Sevilla, 41080 Sevilla, Spain}

\author{N.~Jachowicz}
\affiliation{Department of Physics and Astronomy, Ghent University, B-9000 Gent, Belgium}

\date{\today}

\begin{abstract}
\noindent Single-pion production is an important contribution to the total neutrino-nucleus interaction cross section in accelerator-based neutrino oscillation experiments. The goal of this paper is to improve the Ghent model in the Delta resonance region by incorporating as many physical constraints as possible while keeping the number of fitted parameters as low as possible. A multipole decomposition of the model is performed allowing the use of $K$-matrix theory to unitarize the background contributions. Watson’s theorem is enforced by consistently modifying both the Delta and background contributions across all multipoles. Furthermore, the decay width and form factors of the Delta contribution are modified to ensure compliance with Watson’s theorem, while the model is extended to include $\rho$- and $\omega$-exchange diagrams. These adjustments are compared with other pion production models, as well as with CLAS pion electroproduction data on the nucleon. The results show considerable improvement in the description of the Delta peak region.
\end{abstract}

\maketitle


\section{Introduction}

\noindent The accurate interpretation of accelerator-based neutrino experiments relies strongly on the knowledge of neutrino interaction cross sections~\cite{NuSTEC_white, Katori:2016yel}. In high-energy experiments like DUNE~\cite{Acciarri_DUNE_conceptual_design_report_vol1} and NO$\nu$A~\cite{NOvA_interaction_models_and_uncertainties, NOvA_double_diff_muon_neutrino_cross_sec_measurement} inelastic interactions producing pions are the main contribution to the total neutrino-nucleus cross section. In T2K~\cite{T2K:2011qtm}, Hyper-Kamiokande~\cite{Hyper-K_design_report} and the detectors in the SBN program~\cite{SBN_program_design}, single-pion production (SPP), mainly in the Delta region, accounts for around $20\%$ of the total interaction rate \cite{Isaacson:2025cnk, Hayato:2021heg, Katori:2016yel}. Furthermore, since SPP also contributes to 0-pion final-states, e.g.~when the pion is absorbed or remains otherwise undetected, it constitutes an irreducible background in these experiments. Proper modeling of pion production processes is crucial for both kinematic and calorimetric neutrino energy reconstruction~\cite{Leitner_energy_reconstr, CLAS_energy_reconstr}. A detailed theoretical understanding of SPP interactions is hence pivotal to accurately interpret both the signal and background in accelerator-based neutrino experiments.\\

\noindent Electroweak SPP on the nucleon can be modeled in a systematic way using effective field theory (EFT)~\cite{Dopper:2026qhr, Yao:2018pzc, Yao:2019ngk}. However, the power counting restricts the applicability of the EFT to small
momenta. For an approach that is usable in the large phase space probed in (neutrino) experiments, one typically considers an effective Lagrangian which may include the pions, nucleons, (vector) mesons, and baryon resonances, and is considered at tree-level. Higher order effects are included by parameterizing form factors and by dressing the resonance propagators. Such an approach clearly relies on a large number of degrees of freedom including form factors, resonance widths and relative phases of the different contributions which need to be constrained. As a result, there are a host of different models for electroweak SPP \cite{Hernandez_HNVoffnucleon, Gonzalez-Jimenez:2017fea, Kabirnezhad:2022znc, Lalakulich06}, which include different constraints and have different regions of applicability.\\

\noindent In this paper, improvements to the Ghent hybrid model for SPP~\cite{Raul_Hybrid_model} are presented. This model is based on the approach of Ref.~\cite{Hernandez_HNVoffnucleon}, and includes the Delta resonance and a tree-level background consistent with chiral symmetry breaking. The model was further extended in Refs.~\cite{Hernandez08,Hernandez10,Hernandez13,Alvarez-Ruso_Watson_Delta} with the contribution of higher resonances ($P_{11}(1440)$, $D_{13}(1520)$ and $S_{11}(1535)$), and by imposing unitarity in the Delta-dominated multipoles through Olsson phases~\cite{OLSSON197455}. The Ghent model \cite{Raul_Hybrid_model} extends this model to large invariant masses by making use of a Regge-plus-Resonance approach~\cite{Aznauryan:2003, Aznauryan:2005, DeCruz12a, Corthals06}.
The model has been applied to neutrino and electron induced SPP off nuclei in plane-wave and distorted-wave calculations ~\cite{Garcia-Marcos:2023rnj,Alexis_assessing_paper,Gonzalez-Jimenez:2019qhq, Nikolakopoulos:2018gtf, Gonzalez-Jimenez:2017fea}, and has been embedded in the NuWro event generator~\cite{Yan:2024kkg, Niewczas:2020fev}.\\

\noindent While the main constraints on the model parameters (should) come from data, theoretical considerations such as conservation of vector current (CVC), partial conservation of axial current (PCAC), unitarity and analyticity of
the scattering amplitudes \cite{Berends_dispertion_relations_vol1} can constrain the models further. Unitarity, in particular, is powerful since it can be used to restrict the phases between different contributions to the amplitude. The most complete treatment of unitarity constraints used in neutrino-induced SPP is the Dynamical Coupled Channel (DCC) approach \cite{Doring:2025sgb} of the ANL-Osaka collaboration \cite{DCC_coupled_channel_eq, Kamano:2016bgm, Nakamura:2015rta}.\\

\noindent To include constraints from unitarity in the Ghent model, the $K$-matrix approach \cite{Chung_Kmatrix_theory}, commonly used in unitary isobar models~\cite{Drechsel_MAID2007, Aznauryan:2003, Aznauryan:2005, Drechsel_UIM_up_to_1GeV_(Lagrangians_MAID)}, is applied. Below the two-pion production threshold, the unitarity constraints lead to Watson's theorem~\cite{Watson_Watson's_theorem}. In contrast to the method of Olsson~\cite{OLSSON197455} employed in Ref.~\cite{Alvarez-Ruso_Watson_Delta}, we impose Watson's theorem in both background and resonance contributions. To achieve this, we perform a multipole expansion of the Ghent model. For the direct Delta contribution, we determine an effective-energy dependent width directly from pion-nucleon scattering amplitudes. In addition, we include $\omega$- and $\rho$-exchanges in the vector current, and update the parametrization of the Delta form factors~\cite{Lalakulich06} using the results of the MAID analysis~\cite{Drechsel_MAID2007}. This procedure reduces the number of free parameters in the model, and the reliance on data for electroweak SPP.\\

\noindent In previous work \cite{Alexis_assessing_paper, Raul_Hybrid_model}  we focused on a description of the pion production data over a broad kinematical range. This work aims at refining the fully exclusive description of the Delta region. This region is especially important for experiments like T2K and HK. Also, in the high energy tail of the ESS$\nu$SB beam, pion production in the Delta region is relevant \cite{Alekou:2022emd}. We provide results for the vector current, and compare the results to electron scattering data, and the MAID and DCC analyses. The same procedure can be straightforwardly applied to the axial current; results will be presented in future work.\\

\noindent This work is organized as follows. The kinematics and cross section formula are described in Section \ref{sec: kin and cs}. The Ghent model for SPP is presented in Section \ref{sec: Ghent model}. The unitarity constraints and Watson's theorem are outlined in Section \ref{subsec: unit and Watson}. Incorporating these constraints requires an expansion of the scattering amplitudes in multipoles which is described in \ref{subsec: multi exp}. The implementation of unitarity in the background contributions and the inclusion of the meson exchange diagrams are explained in \ref{sec: opti BG}. The unitarization of the Delta and the parametrization of the Delta form factors are described in \ref{sec: opti res}. The results are shown in Section \ref{sec: results}. In Section \ref{sec: conclusion} we present our conclusion and outlook. Technical details about conventions and calculations are provided in the Appendices.

\section{KINEMATICS AND CROSS SECTION} \label{sec: kin and cs}

\vspace{-0.2cm}

\begin{figure}[H]
    \centering
    \includegraphics[width=\linewidth]{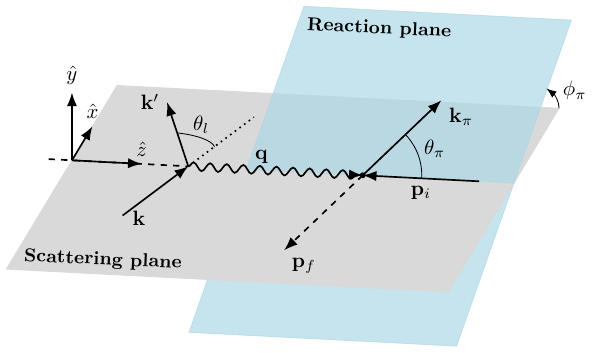}
    \caption{Schematic representation of single-pion production in the CMS.}
    \label{fig: SPP reaction}
\end{figure}

\noindent Single-pion production off the nucleon is shown schematically in Fig.~\ref{fig: SPP reaction}. The incoming lepton with 4-momentum $K^\mu = (E, \bold{k})$ is transformed in the final one with $K'^\mu = (E', \bold{k}')$ by exchanging a single boson, with \mbox{$Q^\mu = K^\mu - K'^\mu = (\omega, \bold{q})$}, with the nucleon. The hadronic part of the interaction may be described in the center of mass system (CMS) defined as

\vspace{-0.4cm}

\begin{align}
    &Q^\mu(\omega, \bold{q}) + P_i^\mu(E_i,p_i = -\bold{q}) \nonumber \\
    = &K_\pi^\mu(E_\pi, \bold{k_\pi}) + P_f^\mu(E_f, p_f = -\bold{k_\pi}).
\end{align}

\noindent It is useful to introduce the Mandelstam variables $s, t$ and $u$ as well as the exchanged 4-momentum squared $Q^2$

\vspace{-0.3cm}

\begin{align}
    &Q^2 = -(K -K')^2, \;s = W^2 = (P_i + Q)^2, \nonumber \\
    &t = (Q - K_\pi)^2, \;u = (P_f - Q)^2,
\end{align}

\noindent where $W$ is the invariant mass. Using these definitions in the CMS yields

\vspace{-0.2 cm}

\begin{align}
    &\omega = \frac{W^2 - M_N^2 - Q^2}{2W},\\
    &E_\pi = \frac{W^2 + m_\pi^2 - M_N^2}{2W}.
\end{align}

\noindent The cross section for SPP can be written as

\vspace{-0.2 cm}

\begin{equation}
    \frac{d^5 \sigma}{dE' d\Omega' d\Omega_\pi} = \frac{F_X^2}{(2\pi)^5}\frac{k'}{E}\frac{k_\pi}{8M_NW}L_{\mu \nu}H^{\mu \nu}, \label{eq: cross section general}
\end{equation}

\noindent where $F_X^2$ is defined as

\begin{align}
    F_{EM}^2 = \frac{1}{2}\left(\frac{4\pi \alpha}{Q^2}\right)^2, \; &F_{CC}^2 = G_F^2 \cos^2(\theta_c), \nonumber \\
    &F_{WNC}^2 = G_F^2,
\end{align}

 \noindent for electromagnetic (EM), charged current (CC) and neutral current (NC) SPP respectively. $\alpha$ is the fine-structure constant, $G_F$ denotes the Fermi-constant and $\theta_C$ stands for the Cabibo-angle. For the magnitude of the momenta, the notation $k \equiv |\bold{k}|$ is used. The lepton tensor is given by

\vspace{-0.2cm}

\begin{equation}
    L_{\mu \nu} = K_\mu K'_\nu + K_\nu K'_\mu - g_{\mu \nu}K_\alpha K'^\alpha - ih\epsilon_{\mu \nu \alpha \beta}K^\alpha K'^\beta.
\end{equation}

\noindent In this expression $h$ is the helicity of the initial lepton. The hadron tensor is defined as

\vspace{-0.2 cm}

\begin{align}
&H^{\mu \nu}(Q^\mu, K_\pi^\mu)\nonumber \\ = &\overline{\sum}_{\lambda_i, \lambda_f}(J^\mu(Q^\mu, K_\pi^\mu, \lambda_i, \lambda_f))^\dagger J^\nu(Q^\nu, K_\pi^\nu, \lambda_i, \lambda_f), \label{eq: hadron tensor}
\end{align}

\noindent where $\overline{\sum}_{\lambda_i, \lambda_f}$ implies a sum over the helicities of the final nucleon $\lambda_f$ and an average over the helicities of the initial nucleon $\lambda_i$. The hadronic current can be written as

\begin{equation}
    J^\mu(Q^\mu, K_\pi^\mu, \lambda_i, \lambda_f) = \bar{u}_{\lambda_f}(\bold{p}_f) \mathcal{O}^\mu_{1\pi}(Q^\mu, K_\pi^\mu)u_{\lambda_i}(\bold{p}_i), \label{eq: hadron current}
\end{equation}

\noindent where the bilinear operator $\mathcal{O}^\mu_{1\pi}(Q^\mu, K_\pi^\mu)$ is defined in Ref.~\cite{Raul_Hybrid_model} and will be discussed later. The helicity spinors $u_{\lambda_i}(\bold{p}_i)$ and $u_{\lambda_f}(\bold{p}_f)$ for initial and final nucleons are defined in Appendix~\ref{sec: App A}.\\

\noindent The cross section for electroproduction of pions is often written as~\cite{Sobczyk:2018ghy}

\vspace{-0.3 cm}

\begin{align}
    \frac{d^5 \sigma}{dE' d\Omega' d\Omega_\pi} = \Gamma_{em}\sum_{i=1}^5 v_i \mathcal{H}_i, \label{eq: cs resp short}
\end{align}

\noindent where

\vspace{-0.4 cm}

\begin{equation}
    \Gamma_{em} = \frac{\alpha}{2\pi^2}\frac{E'}{E}\frac{1}{Q^2}\frac{1}{1-\epsilon}k_\gamma^{LAB},
\end{equation}

\noindent and the longitudinal polarization

\vspace{-0.4 cm}

\begin{equation}
    \epsilon = \left[1 + 2\frac{q_{LAB}^2}{Q^2}\tan^2\left(\frac{\theta_l}{2}\right) \right]^{-1}.
\end{equation}

\noindent The summation in Eq.~(\ref{eq: cs resp short}) can be expressed as

\begin{align}
    \sum_i^5 v_i \mathcal{H}_i = \sigma_0 \Bigg[ &\mathcal{H}_1 + \epsilon \frac{Q^2}{q^2}\mathcal{H}_2\nonumber \\
    + &\sqrt{2\epsilon(1+\epsilon)}\sqrt{\frac{Q^2}{\omega}}\mathcal{H}_3 \cos{\phi_\pi} \nonumber \\
    + &\epsilon \mathcal{H}_4\cos{2\phi_\pi}\nonumber \\
    + &h\sqrt{2\epsilon (1 + \epsilon)}\sqrt{\frac{Q^2} {\omega}}\mathcal{H}_5 \sin{\phi_\pi} \Bigg], \label{eq: hadronic part cross section}
\end{align}

\noindent where $\sigma_0 = \frac{\alpha}{16\pi W^2}\frac{k_\pi}{q}$. The $\mathcal{H}_i$ are defined as

\begin{align}
    &\mathcal{H}_1 = \frac{H^{11} + H^{22}}{2},\label{eq: Hadron 1}\\
    &\mathcal{H}_2 = H^{00},\label{eq: Hadron 2}\\
    &\mathcal{H}_3 = \textrm{Re}(H_{13}),\label{eq: Hadron 3}\\
    &\mathcal{H}_4 = \frac{H_{11} - H_{22}}{2},\label{eq: Hadron 4}\\
    &\mathcal{H}_5 = \textrm{Im}(H_{13}).\label{eq: Hadron 5}
\end{align}

 \noindent The hadronic tensor elements defined in Eqs.~(\ref{eq: Hadron 1}--\ref{eq: Hadron 5}) do not depend on $\phi_\pi$. Consequently, we take $\phi_\pi=0$ when computing these matrix elements. $k_\gamma^{LAB} = \frac{W^2 - M_N^2}{2M_N}$ is the LAB frame momentum of an on-shell photon that would yield the same invariant mass $W$. $q_{LAB} = \sqrt{\omega_{LAB}^2 - Q^2}$ with $\omega_{LAB} = \frac{W^2 - M_N^2 + Q^2}{2M_N}$. Gauge invariance $Q_\mu J^\mu = 0$ is used to write the longitudinal hadron tensor elements in terms of $H^{00}$. After integrating Eq.~(\ref{eq: hadronic part cross section}) over $\phi_\pi$, this expression reduces to

\begin{align}
    &\int_0^{2\pi} d\phi_\pi \sum_i^5 v_i \mathcal{H}_i \nonumber \\
    &= 2\pi \sigma_0 \left(\frac{H_{11} + H_{22}}{2} + \epsilon \frac{Q^2}{q^2}H_{00} \right). \label{eq: cs after int phi}
\end{align}

\noindent The subsequent sections address $\phi_\pi$-integrated cross sections for pion electroproduction, hence only the terms shown in Eq.~(\ref{eq: cs after int phi}) survive, as is the case for non-polarized electron scattering. At the end of this work, the vector-vector contribution to the solid-angles integrated cross section for neutrino induced charged current pion production is discussed.

\begin{figure}[h]

\begin{center}
        \[
\scalebox{0.8}{

\includegraphics[]{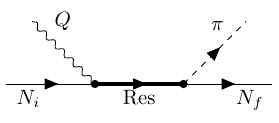}

\qquad

\includegraphics[]{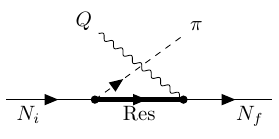}
}
\]
\end{center}
\caption{$s$-channel (left) and $u$-channel (right) diagrams of the resonances in the Ghent model.}
\label{fig: feynman res}
\end{figure}

\begin{figure}[h]

\begin{minipage}[c]{\linewidth}
\begin{center}
    \[
\scalebox{0.8}{

\includegraphics[]{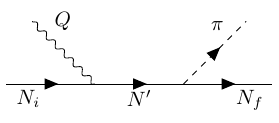}

\qquad

\includegraphics[]{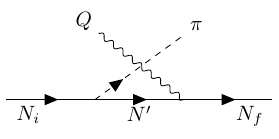}

}
\]
\vspace{-0.5cm}
\[
\scalebox{0.8}{

\includegraphics[]{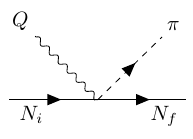}

\quad

\includegraphics[]{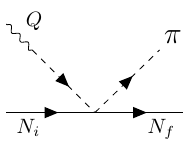}

\quad

\includegraphics[]{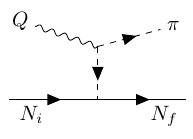}

}
\]
\end{center}
\end{minipage}
    \caption{Background terms computed from the Chiral Perturbation Theory Lagrangian of Ref.~\cite{Hernandez_HNVoffnucleon}. The diagrams in the top row represent the $s$-channel (left) nucleon pole and the $u$-channel (right) crossed nucleon pole. The bottom row shows the contact term (left), pion pole (central) and $t$-channel or pion-in-flight contribution (right).}
    \label{fig: feynman born}
\end{figure}

\section{THE GHENT HYBRID MODEL} \label{sec: Ghent model}

\noindent The starting point of this work is the Ghent model for single-pion production presented in Ref.~\cite{Raul_Hybrid_model}. The Ghent model contains the $s$- and $u$-channel contributions
of the $P_{33}(1232)$ (Delta), $P_{11}(1440)$, $D_{13}(1520)$ and
$S_{11}(1535)$ resonances, shown in Fig.~\ref{fig: feynman res}. In addition, the model contains the background diagrams shown in Fig. \ref{fig: feynman born}, obtained as in Ref.~\cite{Hernandez_HNVoffnucleon}. This background smoothly transitions to a model based on a Reggeized tree level background~\cite{Vanderhaeghen_regge_mesonexchange} around W = 1.5 GeV. The expressions for these contributions to $\mathcal{O}_{1\pi}^ \mu$ as defined in Eq.~(\ref{eq: hadron current}), along with the expression for the transition function, are given in Ref.~\cite{Raul_Hybrid_model}.

\subsection{Unitarity and Watson's theorem} \label{subsec: unit and Watson}

\noindent Watson's theorem \cite{Watson_Watson's_theorem} provides constraints based on unitarity and time reversal invariance. A more in-depth discussion can be found in Ref.~\cite{Alvarez-Ruso_Watson_Delta}. Following the notation of Ref.~\cite{Chung_Kmatrix_theory}, the scattering operator $S$ can be written in terms of the scattering $T$-matrix as

\begin{equation}
    S = \mathbb{1} + 2i \rho^{1/2} T \rho^{1/2} \label{eq: S and T matrix},
\end{equation}

\noindent where $\rho$ is the real and diagonal phase-space matrix making $T$ Lorentz invariant. The dimensionless scattering matrix $\tilde{T}$ is then defined as $\tilde{T} \equiv \rho^{1/2} T \rho^{1/2}$. Consider an initial two-particle helicity state in the CMS $\sket{I}{\Omega_I, \lambda_I}$ where $\Omega_I$ denotes the solid angle and $\lambda_I$ the helicity. The subscript $I$ indicates the channel of the initial state:  $I = \gamma \pi$ or $\pi N$ in the case of photon-induced pion production or pion-nucleon scattering. For a given initial state $I$, the probability $P_F$ of observing the system in an asymptotic state $\sket{F}{\Omega_F, \lambda_F}$ is given by

\begin{equation}
    P_F = |\sbra{F}{\Omega_F,\lambda_F} S \sket{I}{\Omega_I, \lambda_I}|^2,
\end{equation}\\

\noindent assuming the states to be orthonormal and normalized (to one). Since $\sum_F P_F = 1$, $S^\dagger S = \mathbb{1}$ and therefore $S$ is unitary. This implies $i(\tilde{T} - \tilde{T}^\dagger) = -2\tilde{T}^\dagger\tilde{T}$. Taking matrix elements of two-particle CMS helicity states and inserting a complete set of helicity states $\{\sket{\alpha}{\Omega', \lambda'}\}$ yields 

\begin{widetext}

\begin{align}
i\left(\sbra{F}{\Omega_F,\lambda_F}\tilde{T}\sket{I}{\Omega_I,\lambda_I}
- \sbra{F}{\Omega_F,\lambda_F}\tilde{T}^\dagger\sket{I}{\Omega_I,\lambda_I}\right) 
= -2\int d\Omega' \sum_{\lambda',\alpha} 
\sbra{F}{\Omega_F,\lambda_F}\tilde{T}^\dagger\sket{\alpha}{\Omega',\lambda'}
\sbra{\alpha}{\Omega',\lambda'}\tilde{T}\sket{I}{\Omega_I,\lambda_I}. \label{eq: for appendix C}
\end{align}

\noindent Invariance under time reversal leads to $\sbra{F}{\Omega_F,\lambda_F}\tilde{T}^\dagger\sket{I}{\Omega_I,\lambda_I} = \sbra{F}{\Omega_F,\lambda_F}\tilde{T}\sket{I}{\Omega_I,\lambda_I}^*$ \cite{Alvarez-Ruso_Watson_Delta} and as a result the unitarity constraint becomes

\begin{align}
    \int d\Omega' \sum_{\lambda ', \alpha} \sbra{F}{\Omega_F,\lambda_F}\tilde{T}\sket{\alpha}{\Omega',\lambda'}^* \sbra{\alpha}{\Omega',\lambda'}\tilde{T}\sket{I}{\Omega_I,\lambda_I} = -\Im{\sbra{F}{\Omega_F,\lambda_F}\tilde{T}\sket{I}{\Omega_I,\lambda_I}} \in 	\mathbb{R}.\label{eq: unitarity 1}
\end{align}

\end{widetext}

\noindent The diagonal phase-space matrix $\rho$ only adds real factors to $T$ without mixing the matrix elements. Below the two-pion threshold, the only intermediate state that is energetically accessible is the state $\alpha = \pi N$. One can further neglect the product of two electroweak amplitudes since they are suppressed with respect to the strong matrix elements. For SPP, below the two pion production threshold, Eq.~(\ref{eq: unitarity 1}) implies relations between the phases of the $T$-matrix elements for SPP and pion-nucleon scattering:

\begin{align}
    \int d\Omega' \, \sum_{\lambda'} \, &\sbra{\pi N}{\Omega_F,\lambda_F}\tilde{T}\sket{\pi N}{\Omega',\lambda'}^* \nonumber \\
    \times \; &\sbra{\pi N}{\Omega',\lambda'}\tilde{T}\sket{\gamma N}{\Omega_I,\lambda_I} \in \mathbb{R}.\label{eq: Watson}
\end{align}

\noindent In Appendix~\ref{sec: App B}, the helicity amplitudes appearing in Eq.~(\ref{eq: Watson}) are expanded into amplitudes with well-defined total angular momentum $J$. Appendix~\ref{sec: App C} introduces partial-wave amplitudes with fixed pion–nucleon orbital angular momentum $l$ and spin $s$. These amplitudes are combined to construct amplitudes with total isospin $I$ as described in Appendix~\ref{sec: App D}. It is then shown that the constraint of Eq.~(\ref{eq: Watson}) takes a simple form for partial-wave amplitudes with fixed quantum numbers $J, l, s$ and $I$

\begin{align}
    &\sbra{\pi N}{J, l, s, I} \tilde{T}\sket{\pi N}{J, l, s, I}^* \nonumber \\
    \times \; &\sbra{\pi N}{J, l, s, I} \tilde{T}  \sket{\gamma N}{J, L, S, I} \in \mathbb{R}.\label{eq: Watson all QN}
\end{align}

\noindent For the photon–nucleon state, the angular momentum and spin are denoted by $L$ and $S$, as they do not necessarily coincide with the orbital angular momentum $l$ and spin $s$ of the pion–nucleon system. The constraints of Eq.~(\ref{eq: Watson all QN}) are known as Watson's theorem: the phase of each partial-wave amplitude is determined by the strong interaction.

\subsection{Multipole expansion} \label{subsec: multi exp}

\vspace{-0.1 cm}

\noindent In order to be able to implement all the constraints given by Watson's theorem, outlined in Eq.~(\ref{eq: Watson all QN}), the hadronic current is expanded in multipoles. An overview is presented here, together with the connection to the Appendix~\ref{sec: App E}, where a more detailed derivation is provided. The hadronic current given in Eq.~(\ref{eq: hadron current}) can be written in terms of helicity amplitudes. The helicity amplitudes defined in Section \ref{subsec: unit and Watson} in the CMS, as shown in Fig.~\ref{fig: SPP reaction} are defined as

\vspace{-0.5cm}

\begin{align}
    &\epsilon_\mu^r J^\mu(Q^\mu, K_\pi^\mu, \lambda_i, \lambda_f)\nonumber\\
    = &\sbra{\pi N}{\Omega_F, \lambda_F}T\sket{\gamma N}{\Omega_I, \lambda_I}\label{eq: pre helicity for appendix E}\\
    = &\sbra{\pi N}{\Omega_F, \lambda_f}T_r\sket{\gamma N}{\Omega_I, \lambda_i - r},\label{eq: helicity for appendix E}
\end{align}

\vspace{-0.1 cm}

\noindent where $r = -1, 0, +1$ is the helicity of the photon and $\epsilon_\mu^r$ is the polarization vector of the photon as defined in Appendix~\ref{sec: App A}. The helicities of the initial and final nucleon are $\lambda_i$ and $\lambda_f$ respectively. The total helicity of the initial state then becomes $\lambda_I = \lambda_i - r$. Since the pion has no spin, the helicity of the final state is $\lambda_F = \lambda_f$. As shown in Appendix~\ref{sec: App B}, the angular dependence of the helicity amplitudes may be isolated through a decomposition in a series of $J$-projected helicity amplitudes. In the CMS one then has

\vspace{-0.2 cm}

\begin{align}
    &\sbra{\pi N}{\Omega_F, \lambda_F}T_r\sket{\gamma N}{\Omega_I, \lambda_I} = \nonumber \\
    &\sum_{J} \frac{(2J + 1)}{4\pi}\Big(d^J_{\lambda_I, \lambda_F}(\theta_\pi)\Big)^* \sbra{\pi N}{\lambda_F}T_r^J\sket{\gamma N}{\lambda_I}, \label{eq: helicity J amps}
\end{align}

\noindent where $d_{\lambda,\lambda^\prime}^J(\theta)$ are the Wigner small-$d$ matrices containing all the non-trivial angular dependence of the amplitudes. The trivial $\phi_\pi$-dependent phases are discussed in Appendix~\ref{sec: App B}, and are not relevant for the remainder of this work (we choose $\phi_\pi=0$). Note that in the following, a shorter notation $\sbra{\pi N}{J, M, \lambda_F}T\sket{\gamma N}{J, M, \lambda_I} \equiv \bra{\lambda_F}T^J\ket{\lambda_I}$ will be used as $J$ and $M$ are the same for both initial and final state. Also, the subscripts of the initial ($\gamma N$) and final ($\pi N$) states are omitted since they will remain the same. Eqs.~(\ref{eq: pre helicity for appendix E}--\ref{eq: helicity J amps}) define the $J$-projected helicity amplitudes obtained in the model. The behavior of the helicity amplitudes under parity transformation implies that

\begin{equation}
\sbra{}{\lambda_F}T_r^J\sket{}{\lambda_I} = - \sbra{}{-\lambda_F}T_{-r}^J\sket{}{-\lambda_I}.
\end{equation}

\noindent Therefore, Only 4 independent transverse amplitudes $\sbra{}{\lambda_f}T_{r\pm 1}^J\sket{}{\lambda_i \mp 1}$, and 2 independent longitudinal amplitudes $\sbra{}{\lambda_f}T_{r=0}^J\sket{}{\lambda_i}$ remain. To introduce amplitudes with definite final-state pion-nucleon orbital angular momentum $l$, and hence parity $(-1)^{l+1}$, one combines helicity amplitudes with opposite $\lambda_f$ as in Eq.~(\ref{eq: L to J}). One may further define amplitudes with the orbital angular momentum $L$ transferred by the photons. This leads to the well-known transverse electric multipoles $E_{l\pm}$ where $L = l\pm 1$, magnetic multipoles $M_{l\pm}$ where $L=l$, and scalar multipoles $S_{l\pm}$ with $L=l\pm1$. The relation between these multipoles and the $J$-projected helicity amplitudes is derived in Appendix~\ref{sec: App E}, closely following Ref.~\cite{Berends_dispertion_relations_vol1}. This results in

\begin{widetext}
    \begin{align}
    &E_{l+}^V = \frac{-\sqrt{2}}{4i(l+1)}\left(\sqrt{\frac{l}{l+2}}\left(-\bra{\frac{1}{2}}T^{J}_{r=-1}\ket{\frac{3}{2}} - \bra{-\frac{1}{2}}T^{J}_{r=-1}\ket{\frac{3}{2}}\right) +\bra{\frac{1}{2}}T^{J}_{r=-1}\ket{\frac{1}{2}} + \bra{-\frac{1}{2}}T^{J}_{r=-1}\ket{\frac{1}{2}}\right), \label{eq: El+ vec}\\
    &E_{(l+1)-}^V = \frac{-\sqrt{2}}{4i(l+1)}\left(\sqrt{\frac{l+2}{l}}\left(\bra{\frac{1}{2}}T^{J}_{r=-1}\ket{\frac{3}{2}} - \bra{-\frac{1}{2}}T^{J}_{r=-1}\ket{\frac{3}{2}}\right) + \bra{\frac{1}{2}}T^{J}_{r=-1}\ket{\frac{1}{2}} - \bra{-\frac{1}{2}}T^{J}_{r=-1}\ket{\frac{1}{2}}\right), \label{eq: El- vec} \\
    &M_{l+}^V = \frac{-\sqrt{2}}{4i(l+1)}\left(\sqrt{\frac{l+2}{l}}\left(\bra{\frac{1}{2}}T^{J}_{r=-1}\ket{\frac{3}{2}} + \bra{-\frac{1}{2}}T^{J}_{r=-1}\ket{\frac{3}{2}}\right) + \bra{\frac{1}{2}}T^{J}_{r=-1}\ket{\frac{1}{2}} + \bra{-\frac{1}{2}}T^{J}_{r=-1}\ket{\frac{1}{2}}\right), \label{eq: Ml+ vec} \\
    &M_{(l+1)-}^V = \frac{-\sqrt{2}}{4i(l+1)}\left(\sqrt{\frac{l}{l+2}}\left(\bra{\frac{1}{2}}T^{J}_{r=-1}\ket{\frac{3}{2}} - \bra{-\frac{1}{2}}T^{J}_{r=-1}\ket{\frac{3}{2}}\right) - \bra{\frac{1}{2}}T^{J}_{r=-1}\ket{\frac{1}{2}} + \bra{-\frac{1}{2}}T^{J}_{r=-1}\ket{\frac{1}{2}}\right), \label{eq: Ml- vec} \\
    &S_{l+}^V = \frac{-q}{2i(l+1)\sqrt{Q^2}}\left(\bra{\frac{1}{2}}T^{J}_{r=0} \ket{\frac{1}{2}} + \bra{-\frac{1}{2}}T^{J}_{r=0} \ket{\frac{1}{2}}\right), \label{eq: Sl+ vec} \\
    &S_{(l+1)-}^V = \frac{-q}{2i(l+1)\sqrt{Q^2}}\left(\bra{\frac{1}{2}}T^{J}_{r=0} \ket{\frac{1}{2}} - \bra{-\frac{1}{2}}T^{J}_{r=0} \ket{\frac{1}{2}}\right), \label{eq: Sl- vec}
\end{align}

\noindent where the subscript $l\pm$ indicates that the total angular momentum is $J = l \pm \frac{1}{2}$. These multipoles can now be combined to obtain multipoles with a fixed isospin $I$, as explained in Appendix~\ref{sec: App D}. These are the amplitudes of Eq.~(\ref{eq: Watson all QN}) and are then used to compute the elements of the hadron tensor. Expressions are given in Appendix~\ref{sec: App F}. In particular for the angle-integrated hadron tensor elements of Eq.~(\ref{eq: cs after int phi}) one obtains

\begin{align}
    &\int d\Omega_\pi \; \frac{H^{11} + H^{22}}{2} = 2\pi \sum_{l=0} (l+1)^2 \Bigg[ (l+2)(|E_{l+}|^2 + |M_{(l+1)-}|^2) + l(|M_{l+}|^2 + |E_{(l+1)-}|^2) \Bigg],\label{eq: H11 + H22 incl} \\
    &\int d\Omega_\pi \frac{1}{2}H^{00} = 4\pi \sum_{l=0} (l+1)^3 (|S_{l+}|^2 + |S_{(l+1)-}|^2). \label{eq: H00 incl}
\end{align}

\end{widetext}

\noindent These expressions hold for all multipoles (except $E_{0-}, M_{0+}, M_{0-}$ and $E_{1-}$, which are undefined). For all results shown below, all multipoles up to $l=4$ are used as all higher multipoles do not contribute significantly to the responses in the Delta peak region.\\

\noindent This multipole decomposition makes it now possible to implement all the unitarity constraints of Eq.~(\ref{eq: Watson all QN}). In the next Sections, the total $T$-matrix will be separated in a resonant and a background part

\begin{equation}
    T = T_{res} + T_{BG},
\end{equation}

\noindent and both contributions will be discussed separately.

\section{Optimizing the Background}
\label{sec: opti BG}

\noindent First, the modifications of the background are outlined. The background of the Ghent model consists of the tree-level diagrams derived from the HNV Lagrangian shown in Fig.~\ref{fig: feynman born}, as well as the $u$-channel resonance diagrams presented in Fig.~\ref{fig: feynman res}. The latter contributions are treated as background diagrams since the poles are never crossed in real pion production. In addition, $t$-channel meson-exchange diagrams are incorporated, which will be discussed first. The unitarization of the background is then carried out using $K$-matrix theory as described in Sec.~\ref{sec: K matrix theory}.

\subsection{Meson-exchange diagrams} 

\begin{figure}[H]

\[
\scalebox{0.95}{

\includegraphics[]{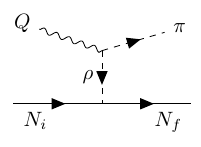}

\qquad

\includegraphics[]{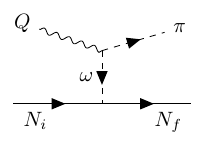}

}
\]
\caption{$t$-channel meson-exchange diagrams for the $\rho$ (left) and $\omega$ meson (right).}
\label{fig: feynman meson}
\end{figure}

\noindent The $t$-channel meson-exchange diagrams added to the model are shown in Fig.~\ref{fig: feynman meson}. The hadronic currents for these contributions are given by Ref.~\cite{Vanderhaeghen_regge_mesonexchange}

\begin{align}
J_\rho^\mu = &\mathcal{I_\rho}e\frac{g_{\rho \pi \gamma}}{m_\pi}g_{\rho NN}F_V(Q^2) \nonumber\\
&\times \Bar{u}_f \epsilon^{\nu \mu \rho \alpha}Q_\nu K^t_\rho \mathcal{P}_{\alpha \beta}\left[\gamma^\beta + \kappa_\rho i \sigma^{\beta \lambda}\frac{K^t_\lambda}{2M_N} \right]u_i, \label{eq: current rho} \\
J_\omega^\mu = &-\mathcal{I_\omega}e\frac{g_{\omega\pi \gamma}}{m_\pi}g_{\omega NN}F_V(Q^2) \nonumber \\
&\times \Bar{u}_f \epsilon^{\nu \mu \rho \alpha}Q_\nu K^t_\rho \mathcal{P}_{\alpha \beta}\left[\gamma^\beta + \kappa_\omega i\sigma^{\beta \lambda}\frac{K^t_\lambda}{2M_N} \right]u_i, \label{eq: current omega}
\end{align}

\noindent with $K^t_\mu = Q_\mu - K_{\pi, \mu}$ the 4-momentum of the exchanged $\rho$ or $\omega$ meson and $\mathcal{P}_{\alpha \beta}$ the spin-1 propagator given by

\begin{equation}
    \mathcal{P}_{\alpha \beta} = \frac{1}{t - m_{mes}^2}\left[-g_{\alpha \beta} + \frac{K^t_\alpha K^t_\beta}{m_{mes}^2} \right],
\end{equation}
 
\noindent where $m_{mes}$ is the mass of the $\rho$ or $\omega$ meson and the $g$-factors denote the couplings of these mesons to the other particles. Further is $\sigma^{\mu \nu} = \frac{i}{2}[\gamma^\mu, \gamma^\nu]$ and the isospin coefficients $\mathcal{I}_{\rho/\omega}$ for the $\rho$- and $\omega$-exchange diagrams are given in Table \ref{tab: isocoef rho}. The values of the coupling constants will be discussed in Sec.~\ref{sec: meson couplings}. In order to provide $Q^2$-dependence for the couplings, an additional dipole form factor

\begin{equation}
    F_V(Q^2) = \frac{1}{\left(1 + \frac{Q^2}{M_V^2}\right)^2} \;,
\end{equation}

\noindent is added with $M_V^2 = 0.71$ GeV$^2$ following the MAID description \cite{Drechsel_UIM_up_to_1GeV_(Lagrangians_MAID)}. 

\begin{table}[H]
    \centering
    \begin{tabular}{|c|c|c|}
\hline
    channel & $\mathcal{I}_\rho$ & $\mathcal{I}_\omega$\\
    \hline
    $\gamma + p \to n + \pi^+$ & $-\sqrt{2}$ & 0\\
    $\gamma + n \to p + \pi^-$ & $-\sqrt{2}$ & 0\\
    $\gamma + p \to p + \pi^0$ & $-1$ & 1\\
    $\gamma + n \to n + \pi^0$ & $1$ & 1\\
    \hline
\end{tabular}
\caption{Isospin coefficients for the $\rho$- and $\omega$-exchange vertices.}
    \label{tab: isocoef rho}
\end{table}

\subsection{Unitarising the background using $K$-matrix theory}\label{sec: K matrix theory}

\noindent To impose Watson's theorem for the background contribution of the $T$-matrix, $K$-matrix theory is used. Using the definition of the $T$-matrix in Eq.~(\ref{eq: S and T matrix}) in terms of the dimensionless $\tilde{T}$, time reversal invariance and unitarity of the $S$-matrix imply

\begin{equation}
    (\tilde{T}^{-1} + i\mathbb{1})^\dagger = (\tilde{T}^{-1} + i\mathbb{1}).
\end{equation}

\noindent This defines a dimensionless real and symmetric matrix containing the same scattering information

\begin{equation}
    \tilde{K}^{-1} \equiv \tilde{T}^{-1} + i\mathbb{1} \Rightarrow \tilde{T} = \tilde{K} + i\tilde{K}\tilde{T}. \label{eq: K from unit}
\end{equation}

\noindent Analogue to $\tilde{T}$, the dimensionless matrix $\tilde{K}$ is defined as $\tilde{K} = \rho^{1/2}K\rho^{1/2}$. Below the two-pion threshold, only two channels should be considered: $\gamma + N$ and $\pi + N$. The phase-space matrix $\rho$ is then given by

\begin{equation}
\rho = 
    \begin{pmatrix}
        \rho_{\gamma} & 0\\
        0 & \rho_{\pi}
    \end{pmatrix}, \label{eq: rho matrix}
\end{equation}

\noindent and the matrices $\tilde{T}$ and $\tilde{K}$ are

\begin{align}
    &\tilde{T} = 
    \begin{pmatrix}
        \sqrt{\rho_{\gamma}} & 0\\
        0 & \sqrt{\rho_{\pi}}\\ 
    \end{pmatrix}
    \begin{pmatrix}
        T_{\gamma \gamma} & T_{\gamma \pi}\\
        T_{\gamma \pi} & T_{\pi \pi}
    \end{pmatrix}
    \begin{pmatrix}
        \sqrt{\rho_{\gamma}} & 0\\
        0 & \sqrt{\rho_{\pi}}\\ 
    \end{pmatrix}\label{eq: T below 2pi}, \\
    &\tilde{K} = 
    \begin{pmatrix}
        \sqrt{\rho_{\gamma}} & 0\\
        0 & \sqrt{\rho_{\pi}}\\ 
    \end{pmatrix}
    \begin{pmatrix}
        K_{\gamma \gamma} & K_{\gamma \pi}\\
        K_{\gamma \pi} & K_{\pi \pi}
    \end{pmatrix}
    \begin{pmatrix}
        \sqrt{\rho_{\gamma}} & 0\\
        0 & \sqrt{\rho_{\pi}}\\ 
    \end{pmatrix}, \label{eq: K below 2pi}
\end{align}

\noindent where $T_{\gamma \gamma}$ stands for the $T$-matrix elements for the reaction $\gamma + N \to \gamma + N$, $T_{\gamma \pi}$ for $\gamma + N \to \pi N$ and $T_{\pi \pi}$ for $\pi + N \to \pi + N$. Inserting these matrices in Eq. (\ref{eq: K from unit}) yields

\begin{align}
    T_{\gamma \pi} = K_{\gamma \pi}(1 + i\rho_\pi T_{\pi \pi}).\label{eq: K matrix unitarization}
\end{align}

\noindent In the derivation of this relation, $T_{\gamma \gamma}$ and $K_{\gamma \gamma}$ have been neglected because these elements are proportional to the square of the EM coupling. The phase-space factor $\rho_\pi$ for pion-nucleon scattering in the CMS is given by \cite{DCC_coupled_channel_eq}

\begin{equation}
    \rho_\pi = \pi \frac{k_\pi E_\pi E_N}{W}. \label{eq: phase-space}
\end{equation}

\noindent The matrices in Eq.~(\ref{eq: K matrix unitarization}) are block-diagonal in $J, l$ and $I$ because of their invariance under rotations in coordinate and isospin space. Because the spin of the pion-nucleon system will always be $s=\frac{1}{2}$, the spin dependence of the $T$- and $K$-matrices is omitted in what follows. As a result, Eq.~(\ref{eq: K matrix unitarization}) can be written for each specific partial wave amplitude with fixed $J$, $l$ and $I$ of the pion-nucleon system

\begin{equation}
    T_{\gamma \pi}^{J, l, I} = K_{\gamma \pi}^{J, l, I}(1 + i\rho_\pi T_{\pi \pi}^{J, l, I}).\label{eq: K matrix unitarization 2}
\end{equation}

\noindent The dimensionless $T$-matrix elements can be written in terms of phase-shifts $\delta_{SPP}$ for SPP and $\delta_{\pi N}$ for pion-nucleon scattering~\cite{Chung_Kmatrix_theory}

\begin{align}
    \Tilde{T}_{\gamma \pi} &= \sin{\delta_{SPP}}e^{i\delta_{SPP}}, \quad \Tilde{T}_{\pi \pi} = \sin{\delta_{\pi N}}e^{i\delta_{\pi N}}.
\end{align}

\noindent Combined with Eq.~(\ref{eq: K matrix unitarization 2}) and $K_{\gamma \pi}$ being a real number, these relations yield

\begin{equation}
    \tan{\delta_{SPP}} = \tan{\delta_{\pi N}},
\end{equation}

\noindent which implies Eq.~(\ref{eq: Watson all QN}).\\

\noindent To unitarize the background contributions, the tree-level $T$-matrix elements are identified with the elements of the $K$-matrix, as motivated in Appendix~\ref{sec: App G}. In the context of dynamical coupled-channel models, these tree-level amplitudes are dressed through a Lipmann-Schwinger equation. Hence, the identification of these tree-level $T$-matrix elements with $K$-matrix elements is equivalent to neglecting the rescattering term in the Lipmann-Schwinger equation~\cite{Kamalov:2000en}. In this work, we use the Julich-Bonn-Washington analysis of Ref.~\cite{Ronchen:2012eg} for the pion-nucleon partial wave amplitudes.\\

\noindent Eq.~(\ref{eq: K matrix unitarization 2}) is also applied above the two-pion threshold. At these higher energies, inelastic pion-nucleon rescattering is possible. This effect is partially taken into account through the inelasticity incorporated in the pion-nucleon scattering amplitudes $T_{\pi N}$. This however, is an approximation. When a new hadronic channel $X$ opens, correction terms proportional to the matrix elements $K_{\gamma X}$ and $K_{\pi X}$~\cite{Workman_K_matrix_multiple_channels} should be added in Eq.~(\ref{eq: K matrix unitarization 2}). These corrections are not taken into account in the current version of the model.

\section{Optimizing the Resonant Part}
\label{sec: opti res}

\noindent Because the focus of this work is on the Delta resonance region, modifications are applied only to this resonance, leaving the higher resonances unchanged. Eq.~(\ref{eq: K matrix unitarization}) is not used directly for unitarizing the Delta contribution since in this case $K_{\gamma\pi}$ has a pole, which is canceled by a zero in $(1+i \tilde{T}_{\pi\pi})$. More details can be found in Appendix~\ref{sec: App H}. Further, the separation between background and resonant contributions is not unique (see Eq.~(\ref{eq: T matrix of delta}) and the discussion below it). Nonetheless, this approach has been used extensively in the extraction of resonance form factors~\cite{Drechsel_MAID2007, Tiator:2011pw, CLAS:2009ces}. In the paragraphs below, the unitarization of the Delta contribution is described. In addition, the form factors are modified using the MAID helicity amplitudes obtained within the MAID-model.

\subsection{Delta resonance and Watson's theorem}

\noindent The helicity amplitudes of the Delta resonance can be written as

\begin{align}
    &\sbra{\pi N}{\Omega_F, \mu_F}T_r^\Delta\sket{\gamma N}{\Omega_I, \mu_I-r}\nonumber \\
    = &\Bar{u}_{\lambda_F} \epsilon_\mu^r \Gamma^\alpha_{\Delta \pi N}S_{\Delta, \alpha \beta} \Gamma^{\mu \beta}_{Q\Delta N}u_{\lambda_I}, \label{eq: helicity of delta}
\end{align}

\noindent where $\Gamma^\alpha_{\Delta \pi N}$ and $\Gamma^{\mu \beta}_{Q\Delta N}$ stand for the $\Delta \pi N$ vertex and $\gamma \Delta N$ vertex respectively. $S_{\Delta, \alpha \beta}$ is the Delta propagator given by

\begin{equation}
    S_{\Delta, \alpha \beta} = \frac{\slashed{k}_\Delta + M_\Delta}{k_\Delta^2 - M_\Delta^2 + iM_\Delta \Gamma_\Delta(W)}\mathcal{P}_{\alpha \beta}, \label{eq: delta propagator}
\end{equation}

\noindent with $\Gamma_\Delta(W)$ the decay width of the Delta and $\mathcal{P}_{\alpha \beta}$ is obtained from Rarita-Schwinger theory \cite{Rarita}

\begin{equation}
    \mathcal{P}_{\alpha \beta} = g_{\alpha \beta} - \frac{1}{3}\gamma_\alpha \gamma_\beta - \frac{2}{3}\frac{k^R_\alpha k^R_\beta}{M_R^2} + \frac{k^R_\alpha \gamma_\beta - k^R_\beta \gamma_\alpha}{3M_R}. \label{eq: rarita}
\end{equation}

\noindent Finally, Eq.~(\ref{eq: helicity of delta}) is multiplied by a cut-off form factor \cite{Raul_Hybrid_model} to regulate the behavior of the Delta at high values of $W$.

\begin{table}[H]
    \centering
\begin{tabular}{|c|l|}
\hline
   $A_4$  & $\ \ \, 3.61395 \cdot 10^{-8}$ MeV$^{-3}$  \\
   $A_3$  & $-1.86647 \cdot 10^{-4}$ MeV$^{-2}$ \\
   $A_2$  & $\ \ \, 0.358868$ MeV$^{-1}$ \\
   $A_1$  & $-3.03853 \cdot 10^{2}$ \\
   $A_0$  & $\ \ \, 9.55320 \cdot 10^4$ MeV \\
    \hline
\end{tabular}

\caption{Fitted constants of Eq. (\ref{eq: fit deltawidth}).}
\label{fig: table fit gamma_delta}
\end{table}

\noindent The Delta resonance contributes to the multipoles with $J = \frac{3}{2}$ and $I = \frac{3}{2}$. As a consequence of Eq.~(\ref{eq: K matrix unitarization}), the background contribution to these multipoles has the correct phase. Eq.~(\ref{eq: Watson all QN}) now demands that the amplitude of Eq.~(\ref{eq: helicity of delta}) also has the same phase as the pion-nucleon scattering amplitude with the same quantum numbers. The phase of the amplitude of Eq.~(\ref{eq: helicity of delta}) is determined by the decay width $\Gamma_\Delta(W)$. This allows to define the decay width as

\begin{equation}
    \Gamma_\Delta(W) = \frac{M_\Delta^2 - s}{M_\Delta}\tan(\delta_{\pi N}(W)), \label{eq: width delta from P33}
\end{equation}

\noindent where $\delta_{\pi N}$ is taken to be the phase shift of the $P_{33}$ partial wave. This method is only valid below the two-pion production threshold. However, for the Delta resonance the $P_{33}$ partial wave remains elastic up to $W \approx 1500$ MeV. As a result, this method is safely extended to these energies. Note that the unitarization of the background contributions combined with this approach for the Delta contribution is consistent with Eq.~(\ref{eq: T matrix of delta}). That is, when $W = M_\Delta$, the contribution of Eq.~(\ref{eq: helicity of delta}) can be identified with $A_{\Delta}$ in (\ref{eq: T matrix of delta}), and the \enquote{background}, which goes to zero at this point, with the $B_\Delta$ term in the same equation.\\

\noindent At $s = M_\Delta^2$ and $\delta_{\pi N} = \pi/2$, the width provided by Eq.\,(\ref{eq: width delta from P33}) is ill-defined. Therefore, at $W = 1232$ MeV the width is put to its experimental value of $120$ MeV \cite{Workman_PDG}. To have a continuous definition of the phase, a polynomial is fitted to Eq.~(\ref{eq: width delta from P33}), 

\begin{align}
    \Gamma_\Delta (W) = A_4 W^4 + A_3 W^3 + A_2 W^2 + A_1 W + A_0,\label{eq: fit deltawidth}
\end{align}

\noindent where the fitted constants are provided in Table~\ref{fig: table fit gamma_delta}.\\

\begin{figure}[H]
    \centering
    \includegraphics[width=\linewidth]{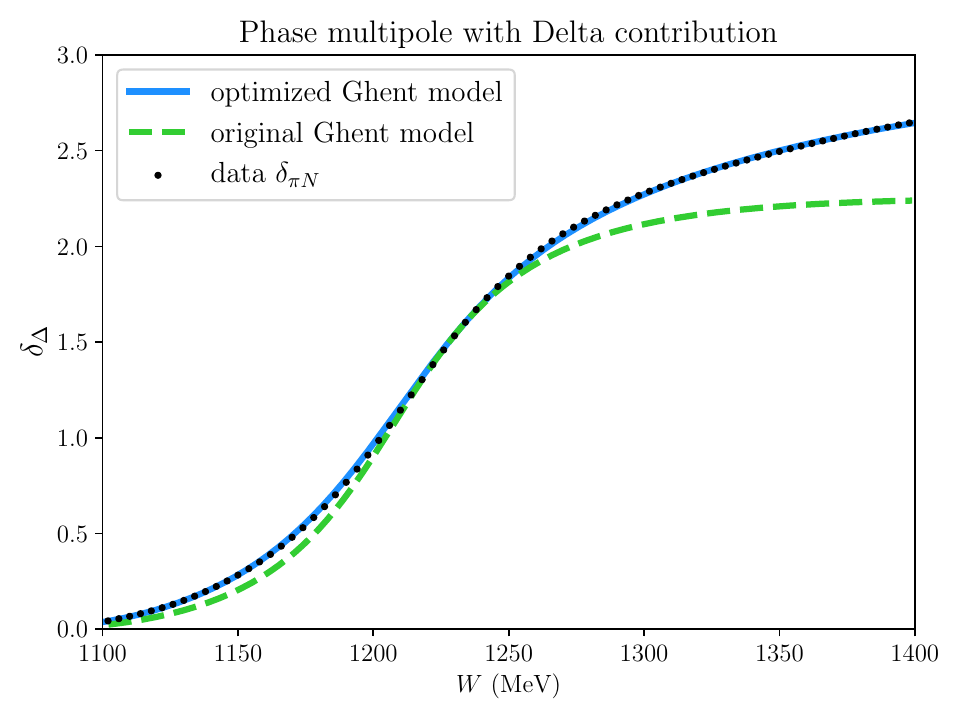}
    \caption{The phase $\delta_{\Delta}$ computed with the fit of $\Gamma_{\Delta}$ of Eq.~(\ref{eq: width delta from P33}) in function of $W$. The green dashed line is the phase obtained with the original Ghent model~\cite{Raul_Hybrid_model}. The data for the pion-nucleon scattering phase is identical to that used for the unitarization of the background diagrams.}
    \label{fig: delta_PIN}
\end{figure}

\noindent In Fig.~\ref{fig: delta_PIN} the pion-nucleon scattering phase determined with Eq.~(\ref{eq: fit deltawidth}) is presented in function of $W$ and compared with the phase of the Delta resonance computed from the decay width used in the original Ghent model \cite{Raul_Hybrid_model}, defined in Ref.~\cite{Leitner_thesis}. Furthermore, the phase from the Julich-Bonn-Washington model used to determine the width is included for comparison. This figure demonstrates that the modified decay width successfully reproduces the phase shift for pion–nucleus scattering across the entire energy range of the Delta peak. The width of the Delta defined in Eq.~(\ref{eq: width delta from P33}) is used up to $W = 1500$ MeV. For larger invariant masses, the width is kept constant at the value at $W=1500$ MeV as the Delta partial wave is no longer elastic and higher mass resonances appear. However, this does not influence the results as the real part of the denominator becomes dominant and the cut-off form factor further suppresses the resonance~\cite{Raul_Hybrid_model}.\\

\noindent It is worthwhile to discuss how our method for unitarization differs from some other approaches.
In Ref~\cite{Alvarez-Ruso_Watson_Delta} an extra $W$- and $Q^2$-dependent phase $e^{i\phi(W,Q^2)}$ is added to the propagator of Eq.~(\ref{eq: delta propagator}). In this case, the background contributions to the $P_{33}$ partial wave are real, and the phase is determined from the pion-nucleon scattering phase-shift such that Eq.~(\ref{eq: Watson all QN}) is satisfied for the Delta dominated $M_{1+}$ multipole. In a $K$-matrix approach on the other hand, the background contribution satisfies Eq.~(\ref{eq: Watson all QN}) for all multipoles. 
In this case, the effective width determined from Eq.~(\ref{eq: width delta from P33}) guarantees that the Delta contribution in the $E_{1+}, M_{1+}$ and $S_{1+}$ multipoles satisfies Eq.~(\ref{eq: Watson all QN}).
This is the unitary isobar model approach of Refs.~\cite{Drechsel_MAID2007, Aznauryan:2003, Aznauryan:2005, Drechsel_UIM_up_to_1GeV_(Lagrangians_MAID)}. The difference between our approach and, e.g., MAID, is that in the latter a specific (Breit-Wigner type) shape for the Delta contribution is assumed.
An additional energy-dependent phase is then added to the resonant contribution in order to satisfy Eq.~(\ref{eq: Watson all QN}). In our case, no initial assumptions on the energy dependence of the Delta self-energy are made.\\

\noindent This approach, however, cannot be applied for the resonances in the second and third resonant region. As explained above, for energies above the two-pion production threshold more decay channels start to open whereby the phase of the resonances is not defined by pion-nucleon scattering only. Additionally, it should be noted that the resonant contribution of Eq.~(\ref{eq: helicity of delta}) contributes not only in the $P_{33}$ partial wave, but also in the $D_{33}$ partial wave. This means that the $D_{33}$ partial waves in principle do not satisfy Eq.~(\ref{eq: Watson all QN}). However, for these partial waves the Delta contribution is small compared to the background. Moreover, the Rarita-Schwinger propagator of Eq.~(\ref{eq: rarita}) is known to not be a pure spin-$3/2$ projector and gives nonphysical contributions in $J=\frac{1}{2}$ partial waves. A way to solve this problem is using consistent couplings that keep only the $J=\frac{3}{2}$ part while projecting out the rest \cite{Pascalutsa:1999zz}. This will be pursued in future work, here we simply removed the direct Delta contributions to $J=\frac{1}{2}$ partial waves.\\

\subsection{Optimizing the Delta form factors}

\noindent The only point where the separation between background and resonance is unambiguous, is at $W= M_{\Delta}$, where the partial wave amplitude is purely imaginary (see Appendix~\ref{sec: App H}). This exact separation is used in unitary isobar models like MAID to define the helicity amplitudes for the production of a specific resonance. For the Delta resonance, these amplitudes are defined in Refs.~\cite{Drechsel_MAID2007, Aznauryan:2008us} and can be linked to the form factors used in the $\Gamma^{\mu \beta}_{Q\Delta N}$-vertex of Eq.~(\ref{eq: helicity of delta}) \cite{Alexis_assessing_paper}:

\begin{widetext}
    \begin{align}
    A_{1/2} &= \sqrt{\frac{\pi \alpha}{3M_N}\frac{(M_\Delta - M_N)^2 + Q^2}{M_\Delta^2 - M_N^2}}\left[\frac{M_N^2 + M_N M_\Delta + Q^2}{M_N M_\Delta}C_3(Q^2) - \frac{Q \cdot K_\Delta}{M_N^2} C_4(Q^2) - \frac{Q\cdot P_i}{M_N^2} C_5(Q^2) \right],\\
    A_{3/2} &= \sqrt{\frac{\pi \alpha}{M_N}\frac{(M_\Delta - M_N)^2 + Q^2}{M_\Delta^2 - M_N^2}}\left[\frac{M_N + M_\Delta}{M_N} C_3(Q^2) + \frac{Q \cdot K_\Delta}{M_N^2}C_4(Q^2) + \frac{Q \cdot P_i}{M_N^2}C_5(Q^2) \right],\\
    S_{1/2} &= \sqrt{\frac{3\pi \alpha}{2M_N}\frac{(M_\Delta - M_N)^2 + Q^2}{M_\Delta^2 - M_N^2}}\frac{q}{M_\Delta} \left[\frac{M_\Delta}{M_N}C_3(Q^2) + \frac{M_\Delta^2}{M_N^2}C_4(Q^2) + \frac{M_\Delta^2 + M_N^2 + Q^2}{2M_N^2}C_5(Q^2) \right].
\end{align}
\end{widetext}

\noindent At the resonance (\mbox{$W = M_\Delta = 1232$ MeV}), these helicity amplitudes can be related to the MAID reduced helicity amplitudes $\Bar{\mathcal{A}}^\Delta_E (M_\Delta, Q^2), \Bar{\mathcal{A}}^\Delta_M(M_\Delta, Q^2)$ and $\Bar{\mathcal{A}}^\Delta_S (M_\Delta, Q^2)$ (see Ref.~\cite{Drechsel_MAID2007} for details)

\begin{align}
    &A_{1/2} + \sqrt{3}A_{3/2} = -2\Bar{\mathcal{A}}^\Delta_M (M_\Delta, Q^2),\\
    &A_{1/2} - \frac{A_{3/2}}{\sqrt{3}} = -2\Bar{\mathcal{A}}^\Delta_E (M_\Delta, Q^2),\\
    &S_{1/2} = -\sqrt{2} \Bar{\mathcal{A}}^\Delta_S (M_\Delta, Q^2).
\end{align}

\noindent We use these relations to formulate the $C_3,C_4$ and $C_5$ form factors in terms of the reduced amplitudes of MAID. Previously, the fit of the Delta form factor to CLAS helicity amplitudes in Ref.~\cite{Lalakulich06} was used. This is updated since, as pointed out in Refs.~\cite{Alexis_assessing_paper, Leitner:2008ue}, the calculation of helicity amplitudes was performed in the lab-frame while data was obtained in the CMS and a different sign convention was used.

\section{Results} \label{sec: results}

\subsection{Inclusive cross sections}

\begin{figure*}[t]
\includegraphics[scale=0.15]{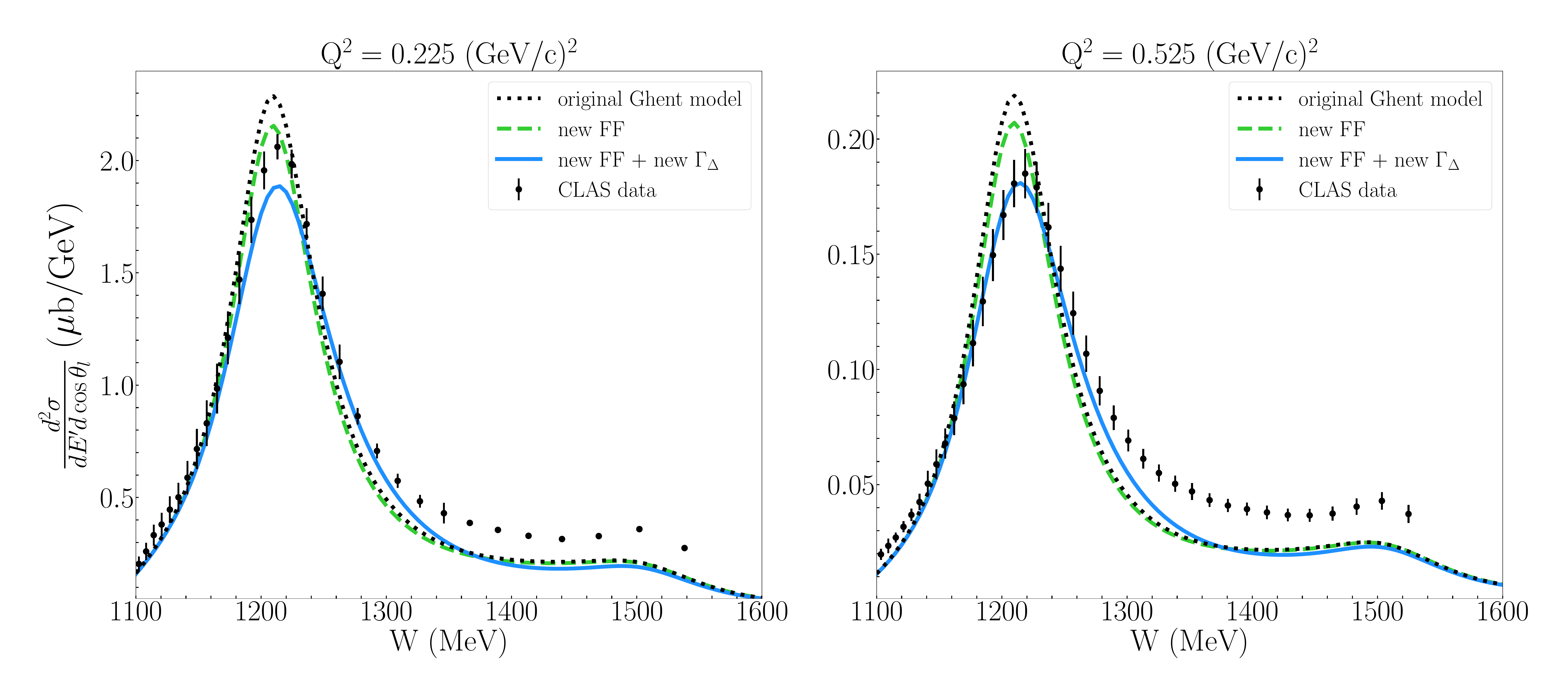}
\caption{Comparing the Ghent model with the original (black dotted) Delta, the Delta with the new form factors (green solid) and the Delta with new form factors and new decay width (blue solid). The $y$-axis shows the inclusive cross section for a fixed incoming electron energy $E = 1.515$ GeV and fixed $Q^2$ values. The inclusive data is taken from the CLAS collaboration \cite{clas_database}.}
\label{fig: delta modifications}
\end{figure*}

\begin{figure*}

\includegraphics[scale=0.15]{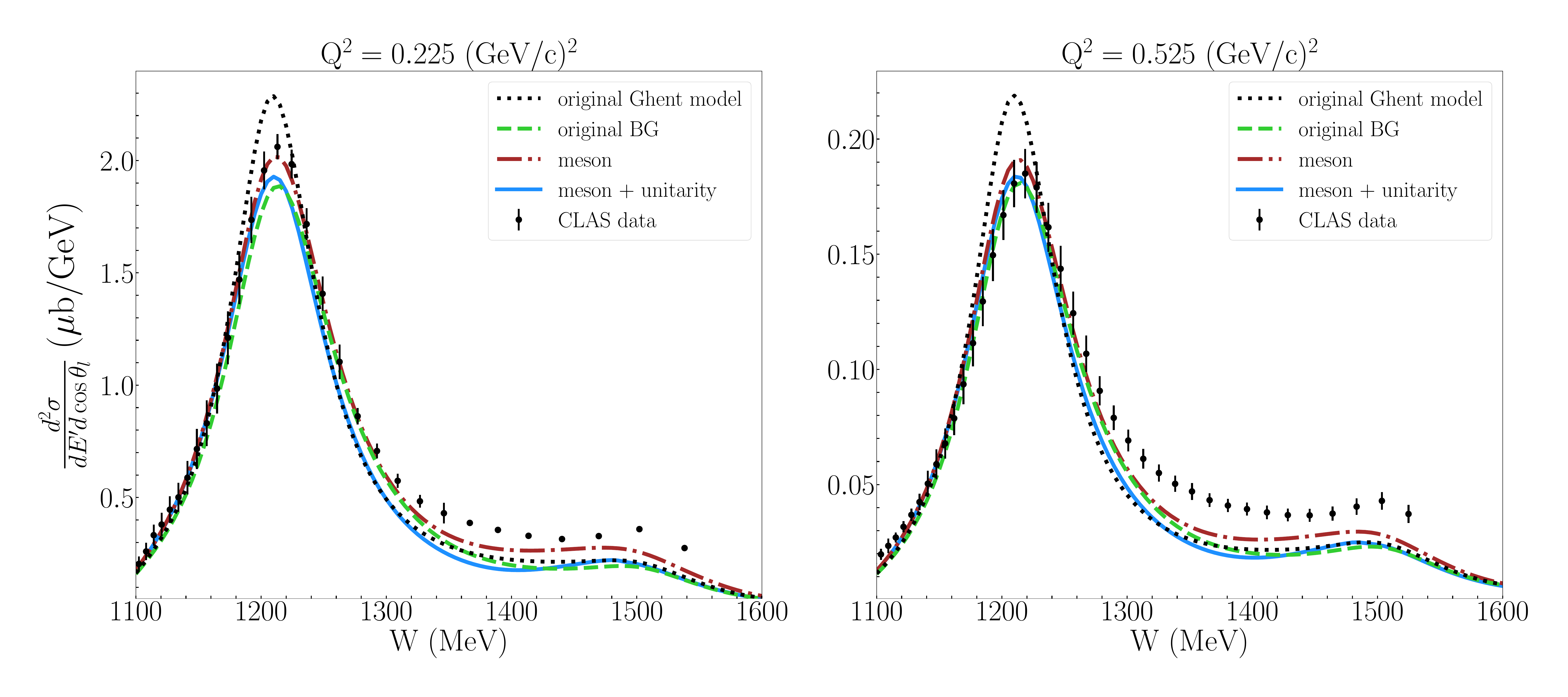}
\caption{Comparing the original background (green dashed), the background with meson contributions (blue dashed) and unitarized background with meson contributions (solid blue) with the modified Delta resonance. This last line is the fully optimized Ghent model. The kinematics is the same as in Fig.~\ref{fig: delta modifications}. The dotted line has no modifications for the Delta or background.}
    \label{fig: background modifications}

\end{figure*}

\noindent In Fig.~\ref{fig: delta modifications}, the modifications of the Delta contribution are presented for inclusive electron-proton scattering cross sections, with an incoming lepton energy of $E = 1.515$ GeV. In these inclusive processes, only the scattered electrons are detected. Since only the one-pion production channel is modeled, the inclusive data should be underpredicted above the two-pion threshold, because only the region up to $W \sim M_{\Delta}$ is fully described by SPP. To isolate the modifications of the Delta resonance, the background is the same as in the previous version of the model \cite{Raul_Hybrid_model} without the unitarizing phases and meson-exchange contributions. In the comparison with the original model, the Olsson-phases introduced in \cite{Alvarez-Ruso_Watson_Delta} are not included. Additionally, data from the CLAS collaboration \cite{clas_database} are included for reference.\\

\noindent The updated form factors reduce the strength of the resonance peak. The modified width leads to an additional reduction of the peak and a small shift to higher $W$ values, in agreement with the data. The low energy tail of the Delta is unaltered by this modification while the peak gets broader at high $W$. These modifications remedy the initial overprediction of the Delta peak. At higher values of $Q^2$, the peak aligns well with the CLAS data, whereas at low momentum transfer it is underestimated. This discrepancy will be solved by the adjustments to the background contributions.\\

\begin{figure*}[!t]
    \centering
    \includegraphics[scale=0.12]{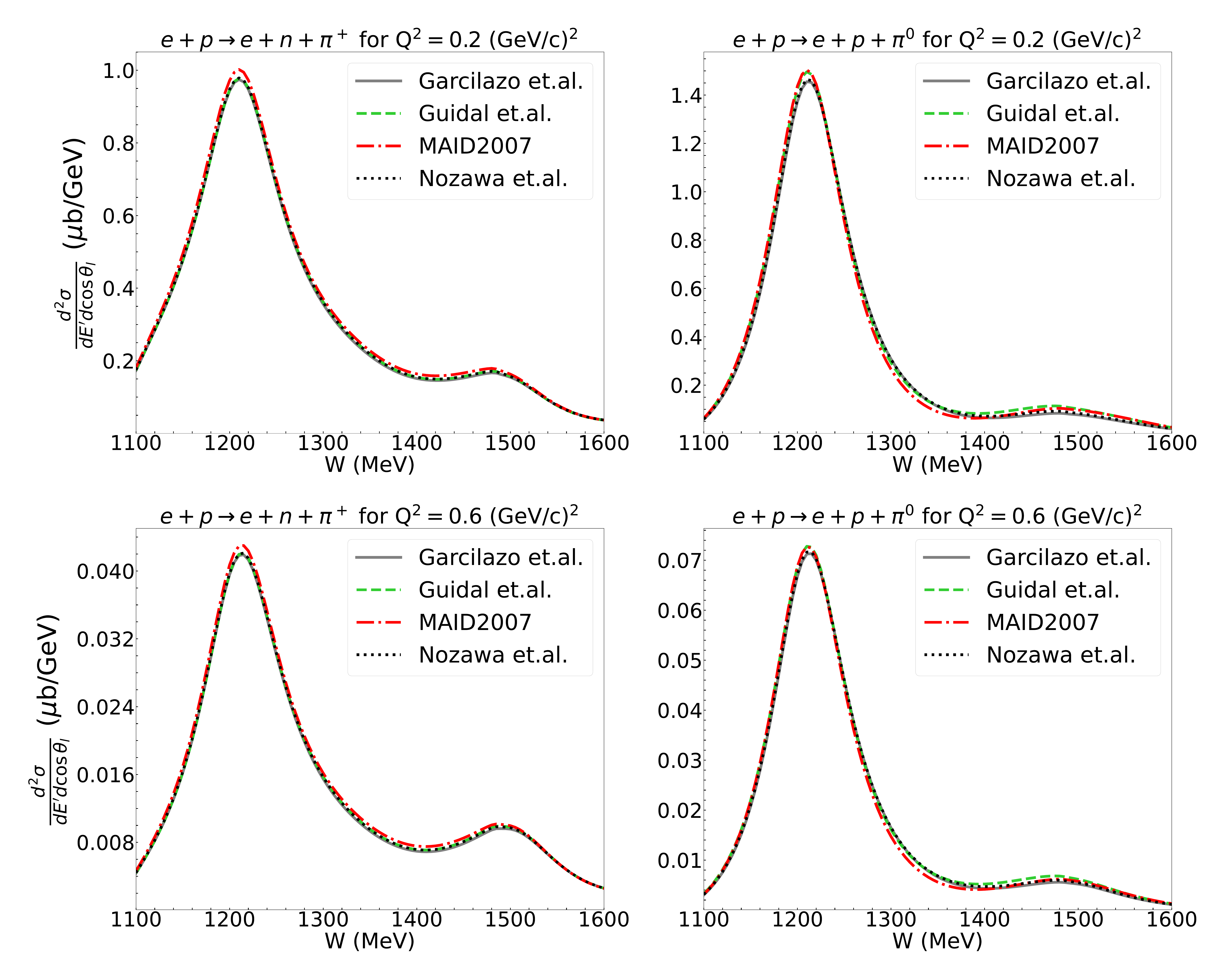}
    \caption{Comparison of the inclusive cross section obtained with different couplings for the $\rho$ and $\omega$ contributions.}
    \label{fig: meson compare}
\end{figure*}

\noindent The effects of adding the $\rho$- and $\omega$-exchanges and unitarizing the background are shown in Fig.~\ref{fig: background modifications}. The kinematics are the same as in Fig.~\ref{fig: delta modifications}. The dotted line shows the model without any modification of the background or Delta resonance. For the curves showing the adjustments, the modified Delta is used. Again, data from CLAS is added as reference. Including the $\rho$ and $\omega$ contributions enhances results in the peak region of the Delta, especially for lower values of $Q^2$. For higher momentum transfer, this reduction is smaller, conserving agreement with the data. Adding the unitarizing factor of Eq.~(\ref{eq: K matrix unitarization}) results in a small reduction over the full $W$ range. This reduction leads to a slight underprediction of the Delta peak at low $Q^2$, while the agreement with the CLAS data at high $Q^2$ is excellent.

\vspace{-0.2 cm}

\subsection{Meson-exchange couplings}\label{sec: meson couplings}

\noindent A substantial amount of work in the literature has investigated the $\rho$- and $\omega$-exchange diagrams in pion production models \cite{Drechsel_MAID2007, Vanderhaeghen_regge_mesonexchange, Garcilazo_rho_omega_param, Nozawa_rho_omega_param}. The currents given in Eqs. (\ref{eq: current rho}) and (\ref{eq: current omega}) are widely employed, and various fits of the coupling constants have been obtained. Tables \ref{fig: table rho couplings} and \ref{fig: table omega couplings} summarize the fitted values obtained by different groups.\\

\noindent Fig.~\ref{fig: meson compare} presents a comparison of the fully modified Ghent model using different sets of coupling constants. The figure demonstrates that the results remain largely consistent across different couplings within the Delta region. Both the Guidal \textit{et al.} and MAID models have a slightly enhanced Delta peak due to their use of stronger couplings compared to other approaches.

\begin{table}[H]
    \centering
    \begin{tabular}{|c|c|c|c|}
\hline
model & $\kappa_\rho$ & $g_{\rho NN}$ & $g_{\rho \pi \gamma}$\\
\hline
H. Garcilazo and E. Moya de Guerra \cite{Garcilazo_rho_omega_param} & 3.71 & 2.66 & 0.103 \\
M. Guidal et.al. \cite{Vanderhaeghen_regge_mesonexchange}& 6.1 & 3.4 & 0.103 \\
S. Nozawa \cite{Nozawa_rho_omega_param} & 3.7 & 2.66 & 0.125 \\
MAID \cite{Drechsel_MAID2007} & 12.7 & 1.8 & 0.103\\
\hline
\end{tabular}
\caption{Coupling constants for the $\rho$-exchange diagrams.}
\label{fig: table rho couplings}
\end{table}

\vspace{0.2 cm}

\begin{table}[H]
\begin{tabular}{|c|c|c|c|}
\hline
model & $\kappa_\omega$ & $g_{\omega NN}$ & $g_{\omega \pi \gamma}$\\
\hline
H. Garcilazo and E. Moya de Guerra \cite{Garcilazo_rho_omega_param} & -0.12 & 7.98 & 0.313  \\
M. Guidal et.al. \cite{Vanderhaeghen_regge_mesonexchange} & 0 & 15 & 0.314 \\
S. Nozawa \cite{Nozawa_rho_omega_param} & 0 & 7.98 & 0.374 \\
MAID \cite{Drechsel_MAID2007} & -0.94 & 16.3 & 0.314\\
\hline
\end{tabular}
\caption{Coupling constants for the $\omega$-exchange diagrams.}
\label{fig: table omega couplings}

\end{table}

\noindent Given that the Delta form factors and $F_V$ added to the meson exchange contributions are determined from MAID, in the following, we use the MAID couplings. As shown in Fig. \ref{fig: meson compare}, the influence of different choices for the coupling constants on the inclusive cross section is minimal.

\vspace{-0.3 cm}

\subsection{Comparison with other models and CLAS data}

\begin{figure*}
    \centering
    \includegraphics[scale=0.16]{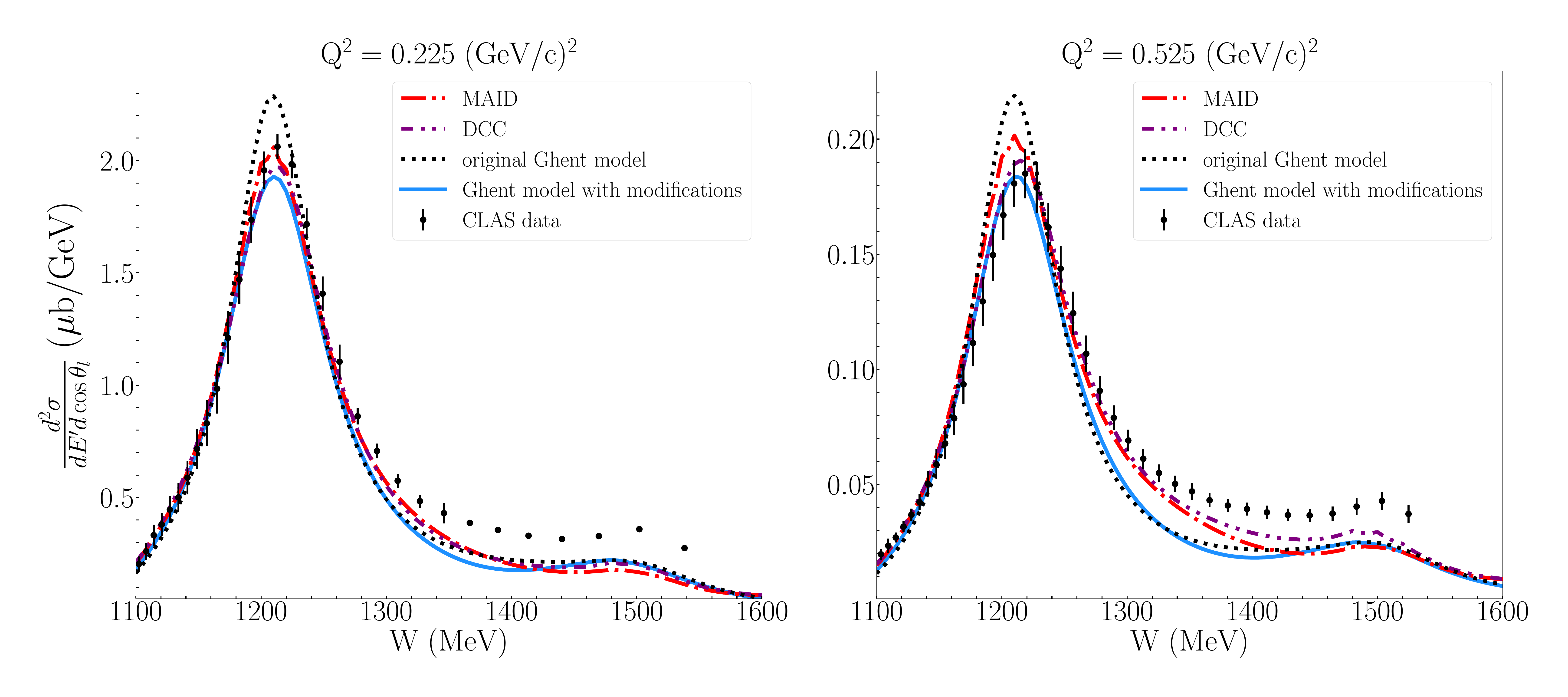}
    \caption{Comparison of the result of the optimized model with data from the CLAS collaboration and the predictions of the MAID model \cite{Drechsel_MAID2007} (red) and DCC model \cite{Matsuyama_DCC_model} (purple). The kinematics are the same as in Fig.~\ref{fig: delta modifications}.}
    \label{fig: inclusive compare models}
\end{figure*}

\noindent In Fig.~\ref{fig: inclusive compare models}, the results for the inclusive cross section of electron-proton scattering after applying all modifications of the Delta and background contributions are compared with the predictions of the MAID \cite{Drechsel_MAID2007} and DCC \cite{Matsuyama_DCC_model} models and with data from the CLAS collaboration.\\

\noindent Fig.~\ref{fig: inclusive compare models} shows that globally the modifications to comply with Watson's theorem improve the agreement with the CLAS data in the Delta peak region. For low values of $Q^2$, the modifications of the Ghent model result in a small underprediction of the height of the Delta peak. However, the predictions are similar to the results of the DCC model. Agreement with data in the peak region is excellent at larger $Q^2$. Both the low- and high-$W$ tail are almost unaltered compared with the original Ghent model. At larger momentum transfer, the high-$W$ tail decreases faster compared to the results of the MAID and DCC models. At these high $W$ values, the data is underpredicted as expected for inclusive electron-proton scattering data. As explained before, beyond the two-pion production threshold the $K$- and $T$-matrices should be extended to account for the new channels that have become energetically accessible. Extension to larger $W$ will be pursued in future work.\\

\noindent Figs.~\ref{fig: exclusive npi+} and \ref{fig: exclusive ppi0} show the exclusive cross section of Eq. (\ref{eq: hadronic part cross section}) where the electrons, final nucleons and pions are detected. These cross sections are integrated over the azimuthal angle $\phi_\pi$ and the results of the MAID and DCC models are shown for comparison. Furthermore, data from the CLAS collaboration \cite{clas_database} for electroproduction of pions on protons is included. Finally, the original Ghent model~\cite{Raul_Hybrid_model} without any modifications and again without including the Olsson phases~\cite{Alvarez-Ruso_Watson_Delta} is added to qualitatively assess the effect of the modifications.\\

\noindent According to Fig.~\ref{fig: exclusive npi+}, it is apparent that after incorporating all the modifications, the Ghent model reproduces the CLAS data well in the case of charged pion production on protons up to $W \approx 1400$ MeV. For more backward pion-angels, the Ghent model tends to slightly overpredict the data above $W = 1400$ MeV. For forward angles, the predictions of the Ghent model lie closer to the results from MAID. At large values of $\theta_\pi$, the results of the Ghent model lie closer with those of the DCC model.

\noindent The results for the neutral pion production channel are shown in Fig.~\ref{fig: exclusive ppi0}. For $W$ up to $1300$ MeV the results of the Ghent model agree with data and the other models, with the exception of the most forward scattering bin, where the data is underpredicted. The high $W$ tail decreases too fast around $\theta_\pi \approx \frac{\pi}{2}$, where the predictions of MAID and the DCC model show better agreement with data in the dip region between the Delta resonance and second resonance region (containing the $P_{11}(1440)$, $D_{13}(1520)$ and $S_{11}(1535)$ resonances). 

\subsection{Vector contribution to charged current neutrino induced pion production}

\noindent The differential cross section for neutrino-induced charged current single pion production in the CMS is given by

\vspace{-0.3 cm}

\begin{equation}
    \frac{d^4 \sigma}{dW dQ^2 d\Omega_\pi} = \frac{G_F^2 \cos^2(\theta_C)}{2(2\pi)^4}\frac{k_\pi}{8M_N^2 E^2}L_{\mu \nu}H^{\mu \nu}.
\end{equation}

\noindent After integration over the pion angles, the double differential cross section becomes

\vspace{-0.3 cm}

\begin{align}
    \frac{d^2\sigma}{dW dQ^2} = &\frac{G_F^2 \cos^2(\theta_C)}{2(2\pi)^3}\frac{k_\pi}{8M_N^2E^2} \Bigg(\frac{L_{00} + L_{22}}{2} (\Tilde{H}^{11} + \Tilde{H}^{22}) \nonumber \\
    + &L_{00} \Tilde{H}^{00} + 2L_{03} \Re (\Tilde{H}^{03}) +L_{33} \Tilde{H}^{33} \nonumber \\
    + &2L_{12}\Im (\Tilde{H}^{12})\Bigg),
\end{align}

\noindent where the tilde on the hadron tensor elements denotes that the integral over pion-angles has been performed. The last contribution is the vector-axial interference term, while the other hadron tensor elements receive both vector-vector and axial-axial contributions. At high energies, the vector-axial interference term is negligible, which is the case for the BEBC experiment~\cite{Alexis_assessing_paper}. We compute the vector-vector contribution to the cross section folded with the BEBC flux \cite{BEBC}. The vector part of the cross section is computed with the same multipoles as in the case of EM SPP. The isospin separation for the CC channel amplitudes is given in Appendix~\ref{sec: App C}.\\

\noindent Fig.~\ref{fig: BEBC plot} shows the vector contribution of the flux-averaged cross section of the Ghent model (with and without modifications) as function of $W$. Vector-vector contributions obtained from the DCC model are added for comparison. For all channels an overall reduction is observed due to the unitarization and modifications of the Delta. The discrepancies between the predictions obtained with the original Ghent model and the DCC model are strongly reduced. In the comparison in Ref.~\cite{Alexis_assessing_paper}, the Olsson phases and a different Regge transition point are used. The fully unitarized model clearly improves agreement with the DCC model, and improves on the results shown in Ref.~\cite{Alexis_assessing_paper}. The vector contribution of the DCC model has been benchmarked against a multitude of data for electromagnetic meson production of both protons and neutrons. Therefore, Fig.~\ref{fig: BEBC plot} indicates that the updated Ghent model will provide more accurate predictions for neutrino-induced pion production.

\section{Conclusions and Outlook} \label{sec: conclusion}

\noindent In this work, we present an updated version of the Ghent model for single-pion production. The aim is to ensure that the predictions in the Delta resonance region fully satisfy Watson’s theorem, which follows from unitarity and time-reversal invariance. Watson's theorem applies to each interaction amplitude with fixed total angular momentum $J$, pion-nucleon orbital angular momentum $l$, spin $s$ and isospin $I$. To incorporate these constraints, a partial wave expansion of the model is performed. The contribution of the Delta resonance is unitarized by adjusting the decay width. In addition, the form factors are updated using the results of the MAID model. The background contributions are extended to include the $\rho$- and $\omega$-exchange diagrams. For the unitarization of the background contributions, $K$-matrix theory is used in which the correct phases are introduced through pion-nucleon scattering amplitudes. In this work, the Julich-Bonn-Washington analysis of Ref.~\cite{Ronchen:2012eg} is used for the latter.\\

\noindent The resulting model is compared with inclusive electron-proton scattering as well as pion electroproduction data off protons from the CLAS collaboration. The results are also compared with the results of the MAID and DCC models. Overall, these comparisons show that the predictions of the Ghent model in the Delta peak region improve considerably for both charged and neutral pion production. The high-energy tail of the Delta remains underpredicted in the neutral pion channel. However, this region lies above the two-pion production threshold where Watson's theorem no longer applies and deviations between data and the model can be expected.\\

\noindent Finally, the effect of these modifications on the vector part of the cross section for CC neutrino-induced single pion production was investigated. The adjustments of the Ghent model decrease the vector part of CC SPP and improve the agreement with the results of the DCC model.\\

\noindent The next step is to extend these modifications to neutrino-induced pion production. For the Delta region, considered here, this is straightforward and results will be presented in an upcoming work. Additionally, work is in progress to incorporate the unitarized model within a nuclear framework, where the initial nucleon is a bound state and the final nucleon and pion are distorted scattering states~\cite{Garcia-Marcos:2023rnj}. Finally, the unitarization scheme should be extended towards higher values of the invariant mass.

\begin{widetext}

\begin{figure}[H]
    \centering
    \includegraphics[scale=0.155]{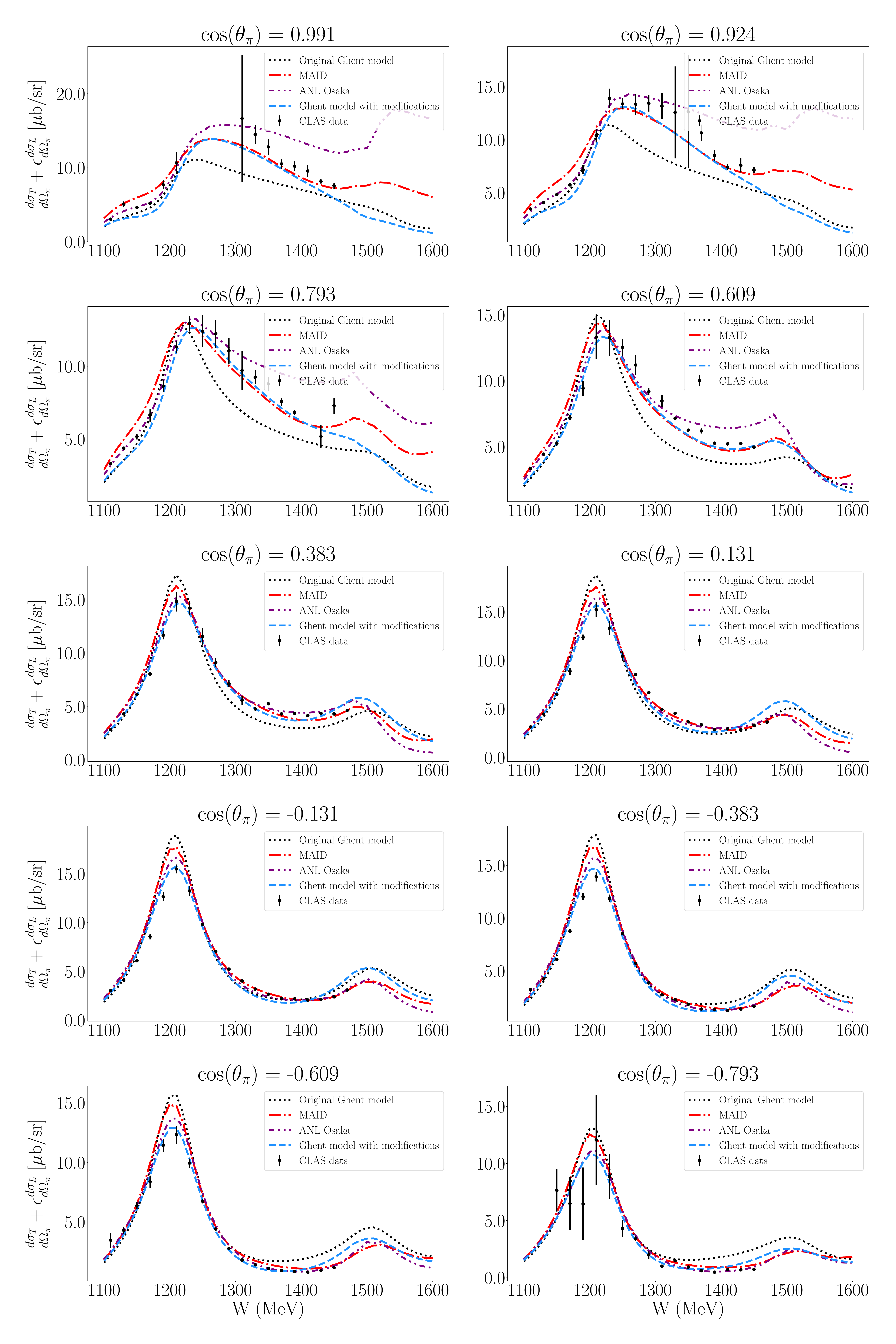}
    \caption{Exclusive cross section for $e + p \to e + n + \pi^+$ at $Q^2 = 0.4$ (GeV/c)$^2$. Here we compare the Ghent model with and without modifications together with CLAS data, DCC and MAID results.}
    \label{fig: exclusive npi+}
\end{figure}

\begin{figure}[H]
    \centering
    \includegraphics[scale=0.155]{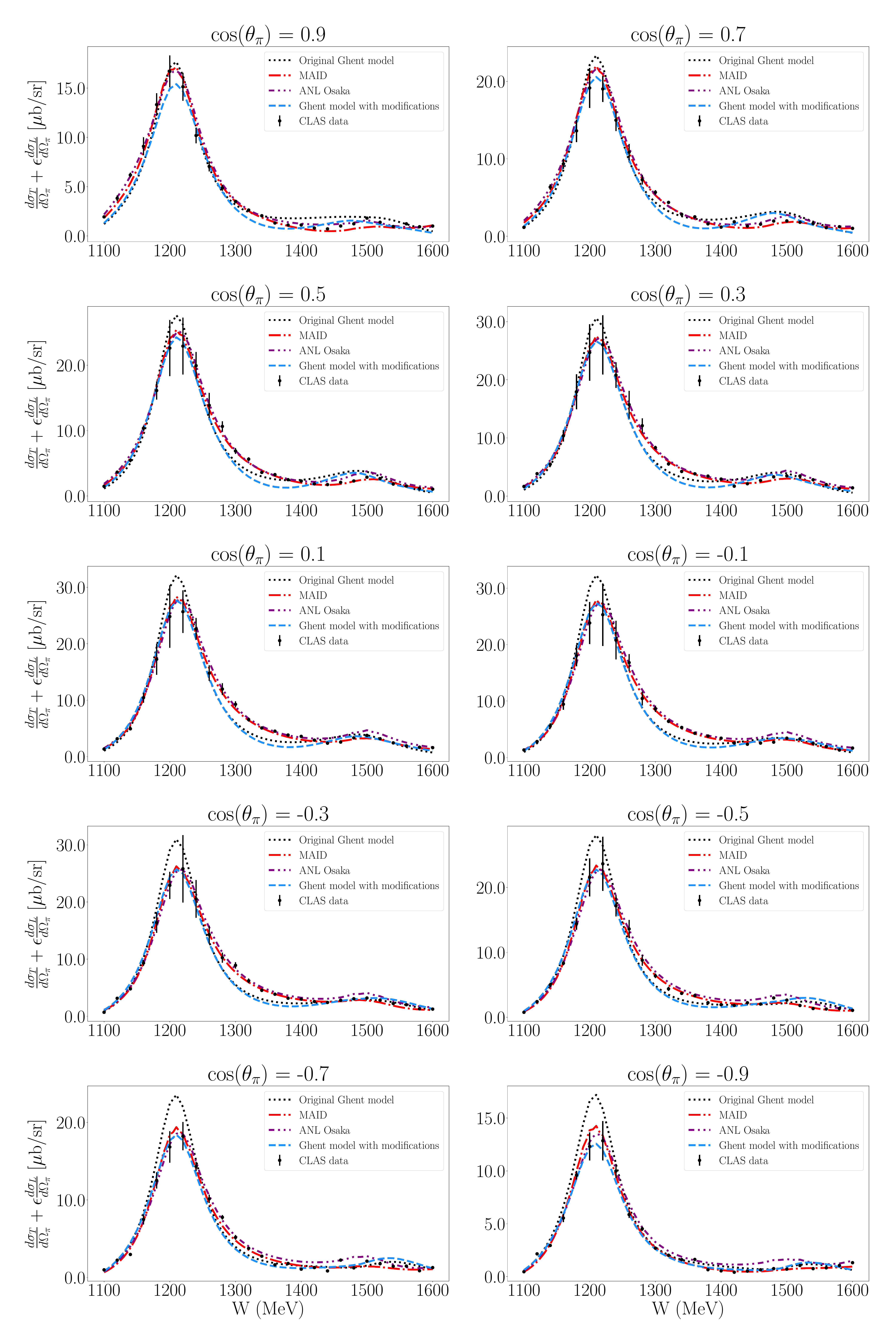}
    \caption{Exclusive cross section for $e + p \to e + p + \pi^0$ at $Q^2 = 0.4$ (GeV/c)$^2$. Here we compare the Ghent model with and without modifications together with CLAS data, DCC and MAID results.}
    \label{fig: exclusive ppi0}
\end{figure}

\begin{figure}[H]
\centering
\includegraphics[scale=0.143]{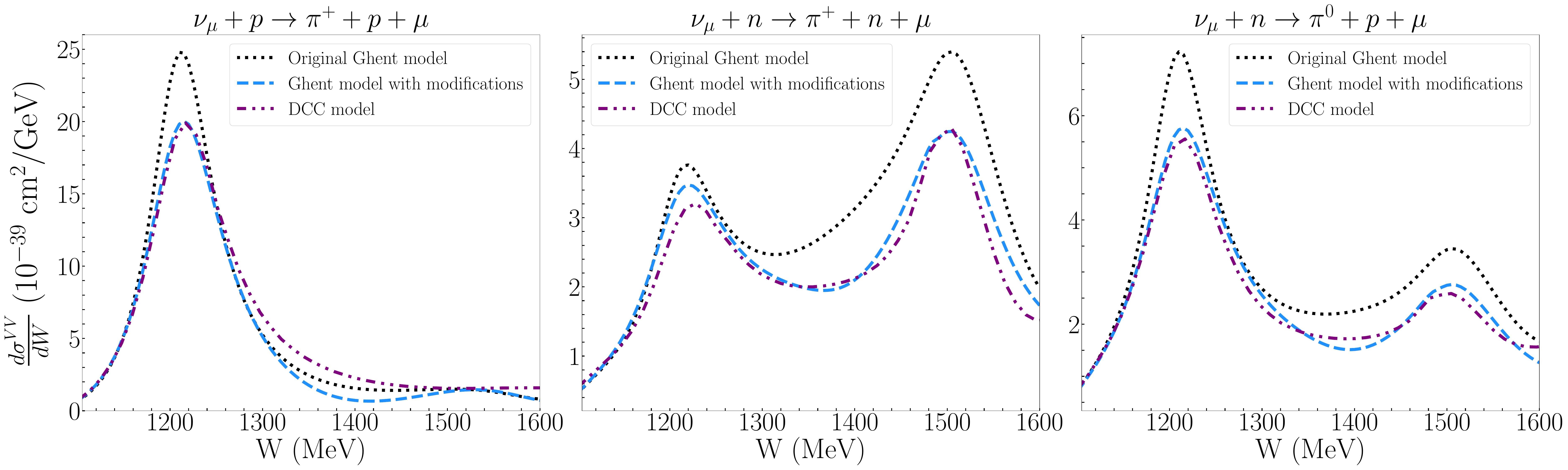}
\caption{$W$-dependence of the vector part of the flux-averaged cross section for CC single pion production using the flux of the BEBC experiment. All cross sections are integrated over $Q^2$ up to $3.5$ (GeV/c)$^2$. For higher values of $Q^2$, the form factors make the cross section negligibly small.}
\label{fig: BEBC plot}
\end{figure}

\end{widetext}

\section*{Acknowledgements}

\noindent R.G.J. was supported by projects PID2021-127098NA-I00 and RYC2022-035203-I funded by MCIN/AEI/10.13039/501100011033/FEDER and FSE+, UE; and by “Ayudas para Atracción de Investigadores con Alto Potencial-modalidad A” funded by VII PPIT-US. J.G.-M. was supported by project PID2021-127098NA-I00 funded by MCIN/AEI/10.13039/501100011033/FEDER; and by the Fund for Scientific
Research Flanders (FWO). A.N. is supported by the Neutrino Theory Network (NTN) under Award Number DEAC02-07CH1135. T. F.-M. was supported by FWO Junior Postdoctoral Fellowship No. 12AG826N. All authors acknowledge support from the FWO and Ghent University Special Research Fund. We thank A. Van Kerckhove for valuable feedback that improved the clarity of the manuscript.

\appendix

\section{Definitions of helicity spinors}\label{sec: App A}

\noindent The four-component helicity spinors used in this work are defined as

\begin{align}
    u_\lambda (\bold{p}) &= u_{\pm}(E,\theta,\phi, \lambda = \pm 1/2) \nonumber \\
    &=
    \begin{pmatrix}
        \sqrt{E+M}\; \chi_\pm (\theta, \phi)\\
        \pm\sqrt{E-M}\; \chi_\pm (\theta, \phi)    \end{pmatrix},
\end{align}

\noindent where $\chi_{\pm}(\theta,\phi)$ are two-component helicity spinors

\begin{align}
    \chi_{+}(\theta, \phi) &= \begin{pmatrix}
        \cos\frac{\theta}{2} \\
        \sin\frac{\theta}{2}e^{i\phi}
    \end{pmatrix}, \\
    \chi_{-}(\theta, \phi) &= \begin{pmatrix}
        -\sin\frac{\theta}{2}e^{-i\phi} \\
        \cos\frac{\theta}{2}
    \end{pmatrix}.
\end{align}

\noindent $\theta \in [0, \pi]$ and $\phi \in [0, 2\pi]$ are the polar and azimuthal angles. In the convention shown in Fig.~\ref{fig: SPP reaction}, the photon momentum is taken to be in the positive $z$-direction. The polar and azimuthal angles of the initial nucleon are then $\theta_i = \pi$ and $\phi_i = 0$. The incoming nucleon helicity spinors in the CMS are

\begin{equation}
    \chi_{i +} = \begin{pmatrix}
        0 \\
        1
    \end{pmatrix},\quad
    \chi_{i -} = \begin{pmatrix}
        -1 \\
        0
    \end{pmatrix}.
\end{equation}

\noindent For the final states, the helicity spinors are

\begin{equation}
    \chi_{f +} = \begin{pmatrix}
        \cos\frac{\theta_f}{2} \\
        -\sin\frac{\theta_f}{2}
    \end{pmatrix},\quad
    \chi_{f -} = \begin{pmatrix}
        \sin\frac{\theta_f}{2} \\
        \cos\frac{\theta_f}{2}
    \end{pmatrix},
\end{equation}

\noindent where the polar angle $\theta_f$ is defined as the angle between the final nucleon and the direction of the photon-momentum. The azimuthal angle of the pion is taken to be $\phi_\pi = 0$  which yields $\phi_f = \phi_\pi + \pi$. It is convenient to write the final spinors in terms of the pion scattering angles $\theta_\pi = \pi - \theta_f$

\begin{equation}
    \chi_{f +} = \begin{pmatrix}
        \sin\frac{\theta_\pi}{2} \\
        -\cos\frac{\theta_\pi}{2}
    \end{pmatrix},\quad
    \chi_{f -} = \begin{pmatrix}
        \cos\frac{\theta_\pi}{2} \\
        \sin\frac{\theta_\pi}{2}
    \end{pmatrix}.
\end{equation}

\noindent The polarization of the exchanged boson can take the values $r = 0, \pm 1$. The transverse polarization vectors are given by $\epsilon^\mu_{r=\pm1} = \mp \frac{1}{\sqrt{2}}(0, 1, \pm i, 0)$. The longitudinal polarization vector is $\epsilon^\mu_{r=0} = \frac{1}{\sqrt{Q^2}}(q, 0, 0, \omega)$ where $\omega$ and $q$ are the energy and 3-momentum of the exchanged boson respectively, as defined in Section \ref{sec: kin and cs}.\\

\section{Total angular momentum states}\label{sec: App B}

\noindent This Appendix shows how helicity amplitudes with fixed total angular momentum J and projection M are  constructed. A more in-depth discussion can be found in Ref.~\cite{wick_wignerD}.\\

\noindent The helicity states defined in Section \ref{sec: Ghent model} obey the orthogonality relation

\begin{align}
  \sbraket{\alpha'}{\Omega', \lambda'}{\Omega, \lambda}{\alpha}
  = \delta^2(\Omega-\Omega')\delta_{\lambda,\lambda'}\delta_{\alpha,\alpha'} ,
\end{align}

\noindent where $\delta^2(\Omega-\Omega') = \delta(\cos(\theta) - \cos(\theta'))\delta(\phi - \phi')$ represents a two-dimensional $\delta$-function on the unit sphere. $\alpha$ and $\alpha'$ are the channels of the states. The helicity states can be expanded in states with fixed angular momentum $J$ and projection on the third axis $M$

\begin{equation}
    \sket{\alpha}{\Omega, \lambda} = \sum_{J, M} \sqrt{\frac{2J+1}{4\pi}}D^J_{M, \lambda}(\Omega)\sket{\alpha}{J, M, \lambda}, \label{eq: hel to J amps}
\end{equation}

\noindent where $D^J_{M, \lambda}(\Omega)$ are the Wigner $D$-matrices. These satisfy the orthogonality relation

\begin{equation}
    \int d\Omega \; (D^{J'}_{M', \lambda}(\Omega))^* D^J_{M, \lambda}(\Omega) = \frac{4\pi}{2J+1}\delta_{J, J'}\delta_{M, M'}. \label{eq: orthogonality WignerD}
\end{equation}

\noindent This relation can be used to write the angular momentum states as

\begin{equation}
    \sket{\alpha}{J, M, \lambda} = \sqrt{\frac{2J+1}{4\pi}}\int d\Omega \; (D^J_{M, \lambda}(\Omega))^* \sket{\alpha}{\Omega, \lambda}. \label{eq: Omega to JM}
\end{equation}

\noindent The overlap between a helicity state and an angular momentum state then becomes

\begin{align}
    &\sbraket{\alpha}{\Omega, \lambda}{J, M, \lambda'}{\alpha'} \nonumber \\
    = &\sqrt{\frac{2J+1}{4\pi}}\int d\Omega' \; (D^J_{M, \lambda'}(\Omega'))^* \sbraket{\alpha}{\Omega, \lambda}{\Omega', \lambda'}{\alpha'}\\
    = &\sqrt{\frac{2J+1}{4\pi}}(D^J_{M, \lambda}(\Omega))^* \delta_{\lambda, \lambda'}\delta_{\alpha, \alpha'}. \label{eq: overlap general}
\end{align}

\noindent The transition operator $T$ is a scalar under rotations, resulting in

\begin{align}
    &\sbra{\alpha'}{J', M', \lambda'}T\sket{\alpha}{J, M, \lambda}\nonumber \\
    = &\sbra{\alpha'}{J, M, \lambda'}T\sket{\alpha}{J, M, \lambda} \delta_{J, J'}\delta_{M, M'}.\label{eq: T scalar}
\end{align}

\noindent Eqs.~(\ref{eq: overlap general}) and (\ref{eq: T scalar}) can be combined to obtain\\

\begin{widetext}

\begin{align}
    &\sbra{F}{\Omega_F, \lambda_F}T\sket{I}{\Omega_I, \lambda_I} \nonumber \\
    &= \sum_{J, M, \lambda, \alpha}\sum_{J', M', \lambda', \alpha'}\sbraket{F}{\Omega_F, \lambda_F}{J, M, \lambda}{{\raisebox{-0.35ex}{$\scriptstyle \alpha$}}} \sbra{\alpha}{J, M, \lambda}T\sket{\alpha'}{J', M', \lambda'} \sbraket{\alpha'}{J', M', \lambda'}{\Omega_I, \lambda_I}{I}\\
    &= \sum_{J, M} \frac{2J+1}{4\pi}(D^J_{M, \lambda_F}(\Omega_F))^* D^J_{M, \lambda_I}(\Omega_I) \sbra{F}{J, M, \lambda_F}T\sket{I}{J, M, \lambda_I}. \label{eq: helicity to J general}
\end{align}

\end{widetext}

\noindent The Wigner $D$-matrices can be written in terms of the Wigner small $d$-matrices $d^J(\theta)$

\begin{equation}
    D^J_{M \lambda}(\Omega) = e^{i(M - \lambda)\phi}d^J_{M \lambda}(\theta).
\end{equation}

\noindent As the azimuthal angle of the final pion can be chosen to be $0$ and the initial particles travel along the $z$-axis such that $\Omega_I = (0, 0)$, Eq.~(\ref{eq: helicity to J general}) becomes

\begin{align}
    \sbra{F}{\Omega_F, \lambda_F}T\sket{I}{\Omega_I, \lambda_I}
    &= \sum_{J, M} \frac{2J+1}{4\pi}d^J_{M \lambda_F}(\theta_F) \nonumber \\
    \times &\sbra{F}{J, M, \lambda_F}T\sket{I}{J, M, \lambda_I}, \label{eq: helicity to J 2}
\end{align}

\noindent where the fact that the Wigner small $d$-matrices are real was used. Notice that this expression differs by a factor $2\pi$ from Ref.~\cite{Berends_dispertion_relations_vol1} as there the integral over $\phi_\pi$ is already performed.

\vspace{0.3 cm}

\section{Unitarity and angular momentum states}\label{sec: App C}

\noindent In this Appendix, Watson's theorem is formulated in terms of amplitudes with fixed $J, l, I$ and parity. The starting point is Eq.~(\ref{eq: for appendix C})

\vspace{-0.2 cm}

\begin{align}
&i\left(\sbra{\pi N}{\Omega_F,\lambda_F}\tilde{T}\sket{\gamma N}{\Omega_I,\lambda_I}
- \sbra{\pi N}{\Omega_F,\lambda_F}\tilde{T}^\dagger\sket{\gamma N}{\Omega_I,\lambda_I}\right) \nonumber \\
= &-2\int d\Omega' \sum_{\alpha, \lambda'} 
\sbra{\pi N}{\Omega_F,\lambda_F}\tilde{T}^\dagger\sket{\alpha}{\Omega',\lambda'} \nonumber \\
&\times \sbra{\alpha}{\Omega',\lambda'}\tilde{T}\sket{\gamma N}{\Omega_I,\lambda_I},
\end{align}

\noindent where $\alpha$ denotes all states that are energetically accessible for a given invariant mass. The tilde can be dropped because the matrix $\rho$ defined in Eq.~(\ref{eq: rho matrix}) is diagonal and only contains real numbers. Inserting the partial wave decomposition of Eq.~(\ref{eq: helicity to J general}) and using time reversal invariance results in

\begin{widetext}

\begin{align}
    &i\sum_{J, M} \frac{2J+1}{4\pi}(D^J_{M, \lambda_F}(\Omega_F))^* D^J_{M, \lambda_I}(\Omega_I) \left( \sbra{\pi N}{J, M, \lambda_F}T\sket{\gamma N}{J, M,\lambda_I} - \sbra{\pi N}{J, M, \lambda_F}T\sket{\gamma N}{J, M, \lambda_I}^* \right) \nonumber \\
    = &-2\sum_{\alpha, \lambda'}\sum_{J, J', M, M'} \int d\Omega' \; \left(\frac{2J+1}{4\pi}\right) \left(\frac{2J' + 1}{4\pi}\right) \left(D^J_{M, \lambda_F}(\Omega_F) \right)^* D^J_{M, \lambda'}(\Omega')\left(D^{J'}_{M', \lambda'}(\Omega') \right)^* D^J_{M', \lambda_I}(\Omega_I) \nonumber \\
    \times &\sbra{\pi N}{J, M, \lambda_F}T\sket{\alpha}{J, M, \lambda'}^* \sbra{\alpha}{J', M', \lambda'}T\sket{\gamma N}{J', M', \lambda_I} \\
    = &-2\sum_{\alpha, \lambda'} \sum_{J, M} \frac{2J + 1}{4 \pi} \left(D^J_{\lambda_I, \lambda_F}(\Omega_F) \right)^* D^J_{M, \lambda_I}(\Omega_I) \sbra{\pi N}{J, M, \lambda_F}T\sket{\alpha}{J, M, \lambda'}^* \sbra{\alpha}{J, M, \lambda'}T \sket{\gamma N}{J, M, \lambda_I},
\end{align}

\noindent where the orthogonality of the Wigner $D$-matrices given by Eq.~(\ref{eq: orthogonality WignerD}) was used to obtain the last expression. Multiplying both sides with $\frac{2J+1}{4\pi}D^J_{M, \lambda_F}(\Omega_F)\left(D^J_{M, \lambda_I}(\Omega_I) \right)^*$ and integrating over $\Omega_F$ and $\Omega_I$ results in

\begin{align}
    &i (\sbra{\pi N}{J, M,\lambda_F}T\sket{\gamma N}{J, M, \lambda_I} - \sbra{\pi N}{J, M, \lambda_F}T\sket{\gamma N}{J, M, \lambda_I}^*) \nonumber \\
    = &-2\sum_{\alpha, \lambda'} \sbra{\pi N}{J, M, \lambda_F}T\sket{\alpha}{J, M, \lambda'}^* \sbra{\alpha}{J, M, \lambda'}T \sket{\gamma N}{J, M, \lambda_I} \in \mathbb{R}.
    \label{eq: unitarity for J amps general}
\end{align}

\noindent This expression provides the constraint from unitarity and time reversal invariance for the $J$-projected helicity amplitudes. It implies relations between the phases of the different amplitudes that are kinematically accessible.\\

\noindent Below the two-pion production threshold and after neglecting the electromagnetic transition $\gamma N \to \gamma N$, only the state $\alpha = \pi N$ remains in the sum over intermediate states. Eq.~(\ref{eq: unitarity for J amps general}) hence reduces to

\begin{align}
    &i (\sbra{\pi N}{J, M,\lambda_F}T\sket{\gamma N}{J, M, \lambda_I} - \sbra{\pi N}{J, M, \lambda_F}T\sket{\gamma N}{J, M, \lambda_I}^*) \nonumber \\
    = &-2\sum_{\lambda'}
\sbra{\pi N}{J, M, \lambda_F}T\sket{\pi N}{J, M, \lambda'}^{*}
\sbra{\pi N}{J, M, \lambda'}T\sket{\gamma N}{J, M, \lambda_I}
\in \mathbb{R}.\label{eq: unitarity for J amps}
\end{align}

\end{widetext}

\noindent This constraint takes a simple form for partial wave amplitudes with total angular momentum $J$, $\pi N$ orbital angular momentum $l$ and spin $s$. Expanding the angular momentum states $\sket{\pi N}{J, M, \lambda}$ of Eq.~(\ref{eq: unitarity for J amps}) in terms of states $\sket{\pi N}{J, M; l, m_l, s, m_s}$ with fixed $l$ and $s$ yields \cite{Watson_Watson's_theorem}

\begin{align}
    \sket{\pi N}{J, M, \lambda} = \sum_{l, m_l, s, m_s} &\bra{J, M; l, m_l, s, m_s}\ket{J, M, \lambda}\nonumber \\
    \times &\sket{\pi N}{J, M; l, m_l, s, m_s}. \label{eq: clebsch LSJ1}
\end{align}

\vspace{0.4 cm}

\noindent The Clebsch-Gordan coefficient in Eq.~(\ref{eq: clebsch LSJ1}) for a two particle state with helicity $\lambda = \lambda_1 - \lambda_2$ in the CMS can be written as

\vspace{-0.5 cm}

\begin{align}
    &\bra{J, M; l, m_l, s, m_s}\ket{J, M, \lambda}\nonumber \\
    = &\sqrt{\frac{2l+1}{2J+1}}\bra{l, m_l, s, m_s}\ket{J, M} \nonumber \\
    &\times \bra{s_\pi, m_{s_\pi}, s_N, m_{s_N}}\ket{s, m_s}\\
    = &\sqrt{\frac{2l+1}{2J+1}}\bra{l, 0, 1/2, \lambda}\ket{J, \lambda} \nonumber \\
    &\times  \bra{0, 0, 1/2, -\lambda_2}\ket{1/2, -\lambda_2}. \label{eq: clebsch l, s, J}
\end{align}

\noindent To obtain Eq.~(\ref{eq: clebsch l, s, J}), we work in the CMS where the quantization of the spins is chosen along the direction of the momentum. In this case the total angular momentum projection $M$ equals the total helicity $M = m_s = \lambda$. As a result the projection of orbital angular momentum $m_l$ is zero. Eq.~(\ref{eq: clebsch LSJ1}) hence simplifies to

\begin{align}
    \sket{\pi N}{J, M, \lambda} =  \sum_{l} &\sqrt{\frac{2l + 1}{2J + 1}} \sbraket{}{l, 0, 1/2, \lambda}{J \lambda}{}\nonumber \\
    \times &\sket{\pi N}{J, M, l, 1/2}.\label{eq: J to L 2}
\end{align}

\noindent This expansion can be inverted

\begin{align}
    \sket{\pi N}{J, M, l, 1/2} = \sum_{\lambda} &\sqrt{\frac{2l + 1}{2J + 1}} \sbraket{}{l, 0, 1/2, \lambda}{J \lambda}{} \nonumber \\
    \times &\sket{\pi N}{J, M, \lambda}.\label{eq: L to J}
\end{align}

\noindent Writing the pion-nucleon scattering amplitude of Eq.~(\ref{eq: unitarity for J amps}) in terms of the $\sket{\pi N}{J, M, l, \frac{1}{2}}$ states, results in

\vspace{0.2 cm}

\begin{widetext}

\begin{align}
    &\sum_{\lambda'} \sbra{\pi N}{J, M, \lambda_F}T\sket{\pi N}{J, M, \lambda'}^* \sbra{\pi N}{J, M, \lambda'}T \sket{\gamma N}{J, M, \lambda_I} \nonumber \\
    = &\sum_{\lambda'}\sum_{l, l'} \sqrt{\frac{2l + 1}{2J + 1}}\sqrt{\frac{2l' + 1}{2J + 1}} \sbraket{}{l, 0, 1/2, \lambda_F}{J, \lambda_F}{}\sbraket{}{J, \lambda'}{l', 0, 1/2, \lambda'}{}  \nonumber \\
    \times &\sbra{\pi N}{J, M, l, 1/2}T\sket{\pi N}{J, M, l', 1/2}^* \sbra{\pi N}{J, M, \lambda'}T\sket{\gamma N}{J, M, \lambda_I} \in \mathbb{R}.
\end{align}

\noindent Due to parity conservation in $\pi N$-scattering, only the $l=l'$ term will yield a non-zero contribution. This implies

\begin{align}
    &\sum_{\lambda'} \sbra{\pi N}{J, M, \lambda_F}T\sket{\pi N}{J, M, \lambda'}^* \sbra{\pi N}{J, M, \lambda'}T \sket{\gamma N}{J, M, \lambda_I} \nonumber \\
    = &\sum_{\lambda'}\sum_{l} \frac{2l + 1}{2J + 1} \sbraket{}{l, 0, 1/2, \lambda_F}{J, \lambda_F}{}\sbraket{}{J, \lambda'}{l, 0, 1/2, \lambda'}{}  \nonumber \\
    \times& \sbra{\pi N}{J, M, l, 1/2}T\sket{\pi N}{J, M, l, 1/2}^* \sbra{\pi N}{J, M, \lambda'}T\sket{\gamma N}{J, M, \lambda_I} \in \mathbb{R}.
\end{align}

\noindent Now Eq.~(\ref{eq: J to L 2}) is used to transform the $\pi N$-state of the second matrix element into a state with fixed $l$ and $s$. This results in

\begin{align}
    &\sum_{\lambda'} \sbra{\pi N}{J, M, \lambda_F}T\sket{\pi N}{J, M, \lambda'}^* \sbra{\pi N}{J, M, \lambda'}T \sket{\gamma N}{J, M, \lambda_I} \nonumber \\
    = &\sum_l \sqrt{\frac{2l + 1}{2J + 1}} \sbraket{}{l, 0, 1/2, \lambda_F}{J, \lambda_F}{}\sbra{\pi N}{J, M, l, 1/2}T\sket{\pi N}{J, M, l, 1/2}^* \sbra{\pi N}{J, M, l, 1/2}T\sket{\gamma N}{J, M, \lambda_I} \in \mathbb{R}.\label{eq: Watson sum L}
\end{align}

\noindent For fixed $J$ the remaining summation only contains the terms with pion-nucleon orbital angular momentum $l = J - \frac{1}{2}$ and $l+1 = J + \frac{1}{2}$. Reducing the notation of the amplitudes to $\sbra{\pi N}{J, M, l, 1/2}T\sket{\pi N}{J, M, l, 1/2} \equiv T_{J,l}^\pi$ and $\sbra{\pi N}{J, M, l, 1/2}T\sket{\gamma N}{J, M, \lambda_I} \equiv T_{J,l}^\gamma$, Eq.~(\ref{eq: Watson sum L}) implies

\begin{align}
    \sqrt{\frac{2l + 1}{2J + 1}}\sbraket{}{l, 0, 1/2, \lambda_F}{J, \lambda_F}{}(T_{J,l}^\pi)^* T_{J,l}^\gamma + \sqrt{\frac{2l+3}{2J + 1}}\sbraket{}{l+1, 0, 1/2, \lambda_F}{J, \lambda_F}{}(T_{J,l+1}^\pi)^* T_{J,l+1}^\gamma \in \mathbb{R}.
\end{align}

\noindent This holds for arbitrary $\lambda_F = \pm \frac{1}{2}$:

\begin{align}
    &\sqrt{\frac{2l + 1}{2J + 1}}\sbraket{}{l, 0, 1/2, 1/2}{J, 1/2}{}(T_{J,l}^\pi)^* T_{J,l}^\gamma + \sqrt{\frac{2l+3}{2J + 1}}\sbraket{}{l+1, 0, 1/2, 1/2}{J, 1/2}{}(T_{J,l+1}^\pi)^* T_{J,l+1}^\gamma \in \mathbb{R},\label{eq: almost lwatson1}\\
    &\sqrt{\frac{2l + 1}{2J + 1}}\sbraket{}{l, 0, 1/2, -1/2}{J, -1/2}{}(T_{J,l}^\pi)^* T_{J,l}^\gamma + \sqrt{\frac{2l+3}{2J + 1}}\sbraket{}{l+1, 0, 1/2, -1/2}{J, -1/2}{}(T_{J,l+1}^\pi)^* T_{J,l+1}^\gamma \in \mathbb{R},\\
    \Rightarrow &\sqrt{\frac{2l + 1}{2J + 1}}\sbraket{}{l, 0, 1/2, 1/2}{J, 1/2}{}(T_{J,l}^\pi)^* T_{J,l}^\gamma - \sqrt{\frac{2l+3}{2J + 1}}\sbraket{}{l+1, 0, 1/2, 1/2}{J, 1/2}{}(T_{J,l+1}^\pi)^* T_{J,l+1}^\gamma \in \mathbb{R},\label{eq: almost lwatson2}
\end{align}

\noindent where the relation for the Clebsch-Gordan coefficients

\begin{align}
    &\sbraket{}{l, m_l, s, m_s}{J, m_J}{}= (-1)^{l + s - J}\sbraket{}{l, -m_l, s, -m_s}{J, -m_J}{},
\end{align}

\noindent was used. Eqs.~(\ref{eq: almost lwatson1}) and (\ref{eq: almost lwatson2}) imply that $(T_{J, l}^\pi)^* T_{J, l}^\gamma \in \mathbb{R}$ for any $l$. This is known as Watson's theorem.

\end{widetext}

\section{Isospin separation of multipoles}\label{sec: App D}

\vspace{-0.1 cm}

\noindent In electroproduction of pions, the multipoles of Eqs.~(\ref{eq: El+ vec}--\ref{eq: Ml- vec}) can be computed for 4 channels:

\vspace{-0.4 cm}

\begin{align}
    &e^- + p \to e^- + \pi^0 + p,\\
    &e^- + p \to e^- + \pi^+ + n,\\
    &e^- + n \to e^- + \pi^0 + n,\\
    &e^- + n \to e^- + \pi^- + p.
\end{align}

\noindent Under the assumption that isospin is a good quantum number, we can write the transition amplitudes between different physical states in terms of isospin amplitudes.
The Wigner-Eckart theorem states
\begin{align}
    &\sbra{F}{I, I_3}T(I^T, I^T_3)\sket{I}{I', I'_3}\nonumber \\
    = &\bra{I', I'_3 ; I^T, I^T_3}\ket{I, I_3}\sbra{F}{I}| T | \sket{I}{I'},\label{eq: wigner-eckart}
\end{align}

\noindent where $I^T$ and $I^T_3$ are the isospin of the $T$-operator and its isospin projection on the third axis. For electroproduction of pions, this operator can be decomposed in an isosvector part $T_1$ with $I^T=1$ and an isoscalar part $T_0$ with $I^T=0$

\vspace{-0.3cm}

\begin{align}
    T^{EM} = T_1(1, I_3^T) + T_0(0, I^T_3).
\end{align}

\noindent Since electromagnetic interactions conserve electric charge, only $I_3^T = 0$ contributes. Pion-nucleon states are now written in terms of total isospin

\begin{align}
    &\sket{\pi N}{I^\pi, I_3^\pi; I_N, I_3^N} \nonumber \\
    = &\sum_I \sum_{I_3 = -I}^I \bra{I^\pi, I_3^\pi; I^N, I_3^N}\ket{I, I_3} \sket{\pi N}{I, I_3}.
\end{align}

\noindent The designation of isospin for the nucleons and pions is listed in Table~\ref{tab: isospin nuc pi}. Denoting

\begin{align}
B_{3/2} &=  \bra{\frac{3}{2}}\bigg| T_1^{EM} \bigg| \ket{\frac{1}{2}}, \\
B_{1/2} &=  \bra{\frac{1}{2}}\bigg| T_1^{EM} \bigg| \ket{\frac{1}{2}}, \\
B^S_{1/2} &=  \bra{\frac{1}{2}}\bigg| T_0^{EM} \bigg| \ket{\frac{1}{2}}, 
\end{align}

\noindent one finds

\vspace{1.4cm}

\begin{align}
     \sbra{\pi^- p}{\lambda_F}T^{EM}\sket{\gamma n}{\lambda_I} &= \nonumber \\
     \frac{\sqrt{2}}{3}&B_{3/2} - \frac{\sqrt{2}}{3}B_{1/2} - \sqrt{\frac{2}{3}}B^S_{1/2} \\
    \sbra{\pi^+ n}{\lambda_F}T^{EM}\sket{\gamma p}{\lambda_I} &= \nonumber \\
    \frac{\sqrt{2}}{3}&B_{3/2} - \frac{\sqrt{2}}{3}B_{1/2} + \sqrt{\frac{2}{3}}B^S_{1/2}\\
    \sbra{\pi^0 p}{\lambda_F}T^{EM}\sket{\gamma p}{\lambda_I} &= \nonumber \\
    \frac{2}{3}&B_{3/2} + \frac{1}{3}B_{1/2} - \sqrt{\frac{1}{3}}B^S_{1/2}\\
    \sbra{\pi^0 n}{\lambda_F}T^{EM}\sket{\gamma n}{\lambda_I} &= \nonumber \\
    \frac{2}{3}&B_{3/2} + \frac{1}{3}B_{1/2} + \sqrt{\frac{1}{3}}B^S_{1/2}.
\end{align}

\noindent Note that we use a different notation compared to the one used in the MAID papers~\cite{Drechsel_UIM_up_to_1GeV_(Lagrangians_MAID), Drechsel_MAID2007}. These relations can be inverted to obtain

\begin{align}
    B^S_{1/2} =& \frac{\sqrt{3}}{2}\Big(\sbra{\pi^0 n}{\lambda_F}T^{EM}\sket{\gamma n}{\lambda_I} \nonumber \\
    &\qquad \qquad - \sbra{\pi^0 p}{\lambda_F}T^{EM}\sket{\gamma p}{\lambda_I}\Big),\label{eq: iso EM 0}\\
    B_{1/2} =& -\frac{1}{2}\sbra{\pi^0 n}{\lambda_F}T^{EM}\sket{\gamma n}{\lambda_I} + \frac{3}{2}\sbra{\pi^0 p}{\lambda_F}T^{EM}\sket{\gamma p}{\lambda_I}\nonumber \\
    &\qquad \qquad \, - \sbra{\pi^- p}{\lambda_F}T^{EM}\sket{\gamma n}{\lambda_I},\label{eq: iso EM 1/2} \\
    B_{3/2} =& \sbra{\pi^0 n}{\lambda_F}T^{EM}\sket{\gamma n}{\lambda_I} \nonumber \\
    & \qquad \qquad +  \frac{1}{\sqrt{2}}\sbra{\pi^- p}{\lambda_F}T^{EM}\sket{\gamma n}{\lambda_I}. \label{eq: iso EM 3/2}
\end{align}

\begin{table}[H]
\begin{center}
    \begin{tabular}{|c|c|c|}
    \hline
      particle & $I$ & $I_3$ \\
      \hline
        $p$ & $1/2$ & $+1/2$ \\
        $n$ & $1/2$ & $-1/2$ \\
        $\pi^+$ & $1$ & $+1$ \\
        $\pi^0$ & $1$ & $0$ \\
        $\pi^-$ & $1$ & $-1$\\
        \hline
    \end{tabular}
    \label{tab: isospin nuc pi}
\end{center}
\caption{Isospin designation of the proton, neutron and pions.}
\end{table}

\noindent For charged-current weak-pion production induced by $W^{\pm}$ bosons, the $T$-operator has a vector and axial-vector contribution

\begin{equation}
T^{CC} = T^{CC}_V(1, I_3^T) + T^{CC}_A(1, I_3^T),
\end{equation}

\noindent where the scalar part is absent since the charged current operator is purely isovector.\\

\noindent The CVC hypothesis now postulates that the vector current in the electromagnetic interaction and the weak interaction are given by components of the same isovector current. As such, we can express the vector amplitudes of the weak charged-current interactions in terms of the same isospin amplitudes

\vspace{-0.2cm}

\begin{align}
    &\sbra{\pi^+ p}{\lambda_F}T^{CC}_V \sket{W^+ p}{\lambda_I} = \sqrt{2}B_{3/2},\\
    &\sbra{\pi^0 p}{\lambda_F}T^{CC}_V \sket{W^+ n}{\lambda_I} = \frac{2}{3}(B_{1/2} - B_{3/2}),\\
    &\sbra{\pi^+ n}{\lambda_F}T^{CC}_V \sket{W^+ n}{\lambda_I} = \frac{\sqrt{2}}{3} (B_{3/2} + 2B_{1/2}).
\end{align}

\noindent These relations can be inverted to obtain

\begin{align}
    B_{3/2} &= \frac{1}{\sqrt{2}} \sbra{\pi^+ p}{\lambda_F}T^{CC}_V \sket{W^+ p}{\lambda_I}, \label{eq: iso CC 3/2} \\
    B_{1/2} &= \frac{3}{2}\sbra{\pi^0 p}{\lambda_F}T^{CC}_V \sket{W^+ n}{\lambda_I} \nonumber \\
    &+ \frac{1}{\sqrt{2}}\sbra{\pi^+ p}{\lambda_F}T^{CC}_V \sket{W^+ p}{\lambda_I}\label{eq: iso CC 1/2}.
\end{align}

\noindent Since total isospin is conserved in the strong interaction, Eq.~(\ref{eq: Watson all QN}) follows for partial wave amplitudes with fixed total isospin given by the linear combinations of Eqs.~(\ref{eq: iso EM 0})-(\ref{eq: iso EM 3/2}) for electromagnetic and Eq.~(\ref{eq: iso CC 3/2}) and (\ref{eq: iso CC 1/2}) for CC pion production.

\vspace{-0.3cm}

\section{From helicity to multipole amplitudes}\label{sec: App E}

\noindent This section explains how the multipole amplitudes in Eqs.~(\ref{eq: El+ vec}--\ref{eq: Sl- vec}) can be obtained starting from the helicity amplitudes given in Eq.~(\ref{eq: helicity for appendix E}). We follow the same procedure and notation as outlined in Ref.~\cite{Berends_dispertion_relations_vol1}.\\

\noindent The pseudo-vector operator in the hadronic currents defined in Eq.~(\ref{eq: hadron current}) can be decomposed in a complete set of $8$ pseudo-vectors $\mathcal{M}_n^\mu$ and associated amplitudes $A_n$, e.g.

\vspace{-0.3cm}

\begin{equation}
    J^\mu = \Bar{u}_f \sum_{n=1}^{8} (A_n \mathcal{M}_n^\mu)u_i.
\end{equation}

\noindent Vector current conservation (CVC) can be used to reduce the amount of independent operators to describe electroproduction of pions to 6. It is convenient to express this current in terms of Pauli-matrices and two-component Pauli spinors with spins along the $z$-axis defined in Appendix~\ref{sec: App A}
\begin{align}
    \sbra{\pi N}{\Omega_F, \mu_F}T_r\sket{\gamma N}{\Omega_I, \mu_I-r} &= \epsilon_\mu^r J^\mu \\
    &= \chi_f^\dagger \epsilon_\mu^r \mathcal{F}^\mu \chi_i, \label{eq: 4vec to 2vec CGLN}
\end{align}

\noindent where the notation for the scattering amplitude is the same as in Eq.~(\ref{eq: helicity for appendix E}). $\epsilon_\mu^r \mathcal{F}^\mu$ stands for the so-called CGLN decomposition given in terms of the CGLN amplitudes $F_i$ \cite{Dennery:1961zz, Chew:1957tf}

\begin{align}
\label{eq: CGLN full}
    \epsilon_\mu^r \mathcal{F}^\mu &= i \boldsymbol{\sigma}\cdot \bold{b^r} F_1 + \frac{\boldsymbol{\sigma}\cdot \bold{k_\pi} \boldsymbol{\sigma}\cdot(\bold{q}\times \bold{b^r)}}{|\bold{k_\pi}||\bold{q}|}F_2 + \frac{i\boldsymbol{ \sigma}\cdot \bold{q} \bold{k_\pi}\cdot \bold{b^r}}{|\bold{k_\pi}||\bold{q}|}F_3 \nonumber \\
    &+ \frac{i\boldsymbol{\sigma}\cdot \bold{k_\pi} \bold{k_\pi}\cdot \bold{b^r}}{|\bold{k_\pi}|^2}F_4 - \frac{i\boldsymbol{\sigma}\cdot \bold{k_\pi}b_0^r}{|\bold{k_\pi}|}F_7 - \frac{i \boldsymbol{\sigma}\cdot \bold{q}b_0^r}{|\bold{q}|}F_8
\end{align}

\noindent where $\epsilon_\mu^r$ is the polarization vector defined in Appendix~\ref{sec: App A} and $b^r_\mu$ is defined as
\begin{equation}
    b_\mu^r = \epsilon_\mu^r - \frac{\boldsymbol{\epsilon^r}\cdot \hat{\bold{q}}}{q}Q_\mu.
\end{equation}

\noindent In Eq.~(\ref{eq: CGLN full}) CVC was used to reduce the amount of independent amplitudes to $6$. After omitting the subscripts of the states and the angular dependencies in Eqs.~(\ref{eq: 4vec to 2vec CGLN}) and using (\ref{eq: CGLN full}), the helicity amplitudes are expressed in terms of the CGLN amplitudes as

\begin{widetext}
    \begin{align}
    &\bra{\frac{1}{2}}T_{r=-1}\ket{\frac{3}{2}} = \chi_{f+}^\dagger (\epsilon_\mu^{r=-1} \mathcal{F}^\mu) \chi_{i +} =  \frac{i}{\sqrt{2}}\sin{ \theta_\pi}\cos{\frac{\theta_\pi}{2}}(F_3 + F_4),\\
    &\bra{-\frac{1}{2}}T_{r=-1}\ket{\frac{3}{2}} = \chi_{f-}^\dagger (\epsilon_\mu^{r=-1} \mathcal{F}^\mu) \chi_{i +} = \frac{i}{\sqrt{2}}\sin{ \theta_\pi}\sin{\frac{\theta_\pi}{2}}(-F_3+F_4),\\
    &\bra{\frac{1}{2}}T_{r=-1}\ket{\frac{1}{2}} = \chi_{f+}^\dagger (\epsilon_\mu^{r=-1} \mathcal{F}^\mu) \chi_{i -} =  \frac{i}{\sqrt{2}}\left(2\cos{\frac{\theta_\pi}{2}}(F_1-F_2) - \sin{ \theta_\pi}\sin{\frac{\theta_\pi}{2}}(F_3-F_4)\right),\\
    &\bra{-\frac{1}{2}}T_{r=-1}\ket{\frac{1}{2}} = \chi_{f-}^\dagger (\epsilon_\mu^{r=-1} \mathcal{F}^\mu) \chi_{i -} = \frac{i}{\sqrt{2}}\left(-2\sin{\frac{\theta_\pi}{2}}(F_1+F_2) - \sin{ \theta_\pi}\cos{\frac{\theta_\pi}{2}}(F_3+F_4)\right).
\end{align}

\noindent The polarization of the boson is chosen to be $r=-1$. The choice $r=+1$ will result in the same multipoles, as there are only 6 independent amplitudes for electroproduction of pions. In matrix notation this system of equations becomes

\begin{align}
    \begin{pmatrix}
        \bra{\frac{1}{2}}T_{r=-1}\ket{\frac{3}{2}} \\
        \bra{-\frac{1}{2}}T_{r=-1}\ket{\frac{3}{2}} \\
        \bra{\frac{1}{2}}T_{r=-1}\ket{\frac{1}{2}} \\
        \bra{-\frac{1}{2}}T_{r=-1}\ket{\frac{1}{2}}
    \end{pmatrix} &= \frac{i}{\sqrt{2}}
    \begin{pmatrix}
        0 & 0 & \sin{ \theta_\pi}\cos{\frac{\theta_\pi}{2}} & \sin{ \theta_\pi}\cos{\frac{\theta_\pi}{2}}\\
        0 & 0 & -\sin{ \theta_\pi}\sin{\frac{\theta_\pi}{2}} & \sin{ \theta_\pi}\sin{\frac{\theta_\pi}{2}}\\
        2\cos{\frac{\theta_\pi}{2}} & -2\cos{\frac{\theta_\pi}{2}} & -\sin{ \theta_\pi}\sin{\frac{\theta_\pi}{2}} & \sin{ \theta_\pi}\sin{\frac{\theta_\pi}{2}}\\
        -2\sin{\frac{\theta_\pi}{2}} & -2\sin{\frac{\theta_\pi}{2}} & -\sin{ \theta_\pi}\cos{\frac{\theta_\pi}{2}} & -\sin{ \theta_\pi}\cos{\frac{\theta_\pi}{2}}
    \end{pmatrix}
    \begin{pmatrix}
        F_1\\
        F_2\\
        F_3\\
        F_4
    \end{pmatrix} = R^{Tr}\begin{pmatrix}
        F_1\\
        F_2\\
        F_3\\
        F_4
    \end{pmatrix}. \label{eq: CGLN to hel tran}
\end{align}

\noindent For the longitudinal part ($r=0$) the same procedure can be followed to obtain

\begin{align}
    &\begin{pmatrix}
        \bra{\frac{1}{2}}T_{r=0}\ket{\frac{1}{2}}\\
        \bra{\frac{1}{2}}T_{r=0}\ket{-\frac{1}{2}}
    \end{pmatrix} = -ib_0^{r=0} \begin{pmatrix}
        \cos{\frac{\theta_\pi}{2}} & \cos{\frac{\theta_\pi}{2}}\\
        \sin{\frac{\theta_\pi}{2}} & -\sin{\frac{\theta_\pi}{2}}
    \end{pmatrix} \begin{pmatrix}
        F_7 \\
        F_8
    \end{pmatrix}= R^{L} \begin{pmatrix}
        F_7 \\
        F_8
    \end{pmatrix}, \label{eq: CGLN to hel long}
\end{align}

\noindent where $b_0^{r=0} = \frac{\sqrt{Q^2}}{q}$. The CGLN amplitudes can be expanded in terms of multipoles \cite{Berends_dispertion_relations_vol1}

\begin{align}
    F_1 = \sum_l &[P'_{l+1}(E_{l+} + lM_{l+}) + P'_{l-1}(E_{l-} + (l+1)M_{l-})], \label{eq: F1}\\
    F_2 = \sum_l &P'_l\left[(l+1)M_{l+} + lM_{l-}\right], \label{eq: F2}\\
    F_3 = \sum_l &\left[P''_{l+1}(E_{l+}-M_{l+}) + P''_{l-1}(E_{l-}+M_{l-})\right],\label{eq: F3}\\
    F_4 = \sum_l &P''_l\left[-E_{l+} - E_{l-} + M_{l+} - M_{l-}\right],\label{eq: F4}\\
    F_7 = \sum_l &P'_l\left[-(l+1) S_{l+} + l S_{l-} \right],\label{eq: F7}\\
    F_8 = \sum_l &\left[(l+1)P'_{l+1}S_{l+} -lP'_{l-1}S_{l-} \right],\label{eq: F8}
\end{align}

\end{widetext}

\noindent where $P_l$ stands for Legendre polynomial with argument $x = \cos{\theta_\pi}$ and the primes denote the derivative of the Legendre polynomials with respect to $x$. These relations can be inverted to obtain

\begin{equation}
\Tilde{M}_l = \int_{-1}^1 dx
\begin{pmatrix}
D_l(x) & 0\\
0 & E_l(x)
\end{pmatrix} \Tilde{F}, \label{eq: CGLN to multi}
\end{equation}

\vspace{0.5 cm}

\noindent where the matrices $\Tilde{M}_l$, $\Tilde{F}$, $D_l(x)$ and $E_l(x)$ are given by

\begin{equation}
    \Tilde{M}_l = 
    \begin{pmatrix}
    E_{l+}\\
    E_{l-}\\
    M_{l+}\\
    M_{l-}\\
    S_{l+}\\
    S_{l-}
    \end{pmatrix}, \qquad \Tilde{F} = 
    \begin{pmatrix}
        F_1\\
        F_2\\
        F_3\\
        F_4\\
        F_7\\
        F_8
    \end{pmatrix},
\end{equation}

\begin{align}
    &D_l(x) = \nonumber \\
    &\begin{pmatrix}
    \frac{P_l}{2(l+1)}&-\frac{P_{l+1}}{2(l+1)}&\frac{l(P_{l-1}-P_{l+1})}{2(2l+1)(l+1)}&\frac{P_l - P_{l+2}}{2(2l+3)}\\
    \frac{P_l}{2l}&-\frac{P_{l-1}}{2l}&\frac{(l+1)(P_{l+1}-P_{l-1})}{2l(2l+1)}&\frac{P_l-P_{l-2}}{2(2l-1)}\\
    \frac{P_l}{2(l+1)}&-\frac{P_{l+1}}{2(l+1)}&\frac{P_{l+1}-P_{l-1}}{2(l+1)(2l+1)}&0\\
    -\frac{p_l}{2l}&\frac{P_{l-1}}{2l}&\frac{P_{l-1}-P_{l+1}}{2l(2l+1)}&0
    \end{pmatrix},
\end{align}

\vspace{0.5 cm}

\noindent and

\vspace{0.5 cm}

\begin{equation}
    E_l(x) = \frac{1}{2}
    \begin{pmatrix}
    \frac{P_{l+1}}{l+1} & \frac{P_l}{l+1}\\
    \frac{P_{l-1}}{l} & \frac{P_l}{l}
    \end{pmatrix}.
\end{equation}

\vspace{0.5 cm}

\noindent Eqs. (\ref{eq: CGLN to hel tran}), (\ref{eq: CGLN to hel long}) and (\ref{eq: CGLN to multi}) can be used to obtain

\begin{align}
    \Tilde{M}_l = \int_{-1}^1 dx
&\begin{pmatrix}
D_l(x) & 0\\
0 & E_l(x)
\end{pmatrix}\nonumber \\
\times 
&\begin{pmatrix}
    (R^{Tr})^{-1} & 0\\
    0 & (R^{L})^{-1}
\end{pmatrix} \Tilde{T},
\end{align}

\noindent with

\begin{equation}
    \Tilde{T} = \begin{pmatrix}
        \bra{\frac{1}{2}}T_{r=-1}\ket{\frac{3}{2}} \\
        \bra{-\frac{1}{2}}T_{r=-1}\ket{\frac{3}{2}} \\
        \bra{\frac{1}{2}}T_{r=-1}\ket{\frac{1}{2}} \\
        \bra{-\frac{1}{2}}T_{r=-1}\ket{\frac{1}{2}} \\
        \bra{\frac{1}{2}}T_{r=0}\ket{\frac{1}{2}} \\
        \bra{\frac{1}{2}}T_{r=0}\ket{-\frac{1}{2}}
    \end{pmatrix}.
\end{equation}

\noindent This expression can be written in terms of helicity amplitudes with a fixed total angular momentum $J = l' + \frac{1}{2}$ using Eq.~(\ref{eq: helicity to J 2})

\begin{align}
     \Tilde{M}_l = &\int_{-1}^1 dx
\begin{pmatrix}
D_l(x) & 0\\
0 & E_l(x)
\end{pmatrix}
\begin{pmatrix}
    (R^{Tr})^{-1} & 0\\
    0 & (R^{L})^{-1}
\end{pmatrix} \nonumber \\
\times &\sum_{l'}\Tilde{O}_{l'}\Tilde{T}^{l'},
\end{align}

\noindent where

\begin{equation}
    \Tilde{T}^{l'} = \begin{pmatrix}
    \bra{\frac{1}{2}}T^{l'}_{r=-1}\ket{\frac{3}{2}}\\
    \bra{-\frac{1}{2}}T^{l'}_{r=-1}\ket{\frac{3}{2}}\\
    \bra{\frac{1}{2}}T^{l'}_{r=-1}\ket{\frac{1}{2}}\\
    \bra{-\frac{1}{2}}T^{l'}_{r=-1}\ket{\frac{1}{2}}\\
    \bra{\frac{1}{2}}T^{l'}_{r=0}\ket{\frac{1}{2}}\\
    \bra{\frac{1}{2}}T^{l'}_{r=0}\ket{-\frac{1}{2}}
    \end{pmatrix},
\end{equation}

\noindent and $\Tilde{O}_{l'}$ contains the Wigner small $d$-matrices of Eq.~(\ref{eq: helicity to J 2}). These  matrices can be written in terms of Legendre polynomials \cite{wick_wignerD}. $\Tilde{O}_{l'}$ then becomes the diagonal matrix

\begin{align}
    &\Tilde{O}_{l'} = \nonumber \\
    &diag\Bigg(-\sin{\frac{\theta_\pi}{2}}\left[\sqrt{\frac{l'}{l'+2}}P'_{l'+1}+\sqrt{\frac{l'+2}{l'}}P'_{l'}\right], \nonumber \\
    &\cos{\frac{\theta_\pi}{2}}\left[-\sqrt{\frac{l'}{l'+2}}P'_{l'+1}+\sqrt{\frac{l'+2}{l'}}P'_{l'}\right], \nonumber\\ &\cos{\frac{\theta_\pi}{2}}\left[P'_{l'+1}-P'_{l'}\right], \; \sin{\frac{\theta_\pi}{2}}\left[P'_{l'+1}+P'_{l'}\right], \nonumber \\
    &\cos{\frac{\theta_\pi}{2}}\left[P'_{l'+1}-P'_{l'}\right], \; \sin{\frac{\theta_\pi}{2}}\left[P'_{l'+1}+P'_{l'}\right]\Bigg).
\end{align}

\noindent After using orthogonality relations for the Legendre polynomials and some straighforward manipulations, one arrives at the expressions given in Eqs.~(\ref{eq: El+ vec}--\ref{eq: Sl- vec}).

\vspace{-0.4 cm}

\section{Responses in terms of multipoles}\label{sec: App F}

\noindent For electroproduction of pions on a nucleon, the only terms that survive in the cross section after integrating over $\phi_\pi$ are given in Eq.~(\ref{eq: cs after int phi}). In this section, the calculations for these hadronic tensor elements in terms of multipoles will be sketched.\\

\noindent The hadronic currents $J^0, J^1$ and $J^2$ can be written in terms of CGLN amplitudes using Eq. (\ref{eq: 4vec to 2vec CGLN})

\vspace{-0.4cm}

\begin{align}
    J^0 = &\chi_f^\dagger[i\sigma_3 (\cos{\theta_\pi}F_7 + F_8) + i\sigma_1\sin{ \theta_\pi}F_7]\chi_i,\\
    J^1 = &\chi_f^\dagger[i\sigma_1(F_1 - \cos{\theta_\pi}F_2 + \sin^2{\theta_\pi}F_4)\nonumber \\
    &+ i\sigma_3 \sin{ \theta_\pi}(F_2 + F_3 + \cos{\theta_\pi}F_4 ]\chi_i,\\
    J^2 = &\chi_f^\dagger\left[i\sigma_2(F_1 - \cos{\theta_\pi}F_2) -\mathbb{1}\sin{ \theta_\pi}F_2 \right]\chi_i.
\end{align}

\noindent From these currents, one obtains the hadron tensor elements

\vspace{-0.4cm}

\begin{align}
    \frac{1}{2}&\sum_{\lambda_i, \lambda_f} \frac{H^{11} + H^{22}}{2} = |F_1|^2 + |F_2|^2 \nonumber \\ 
    &+ \frac{1}{2}\sin^2{\theta_\pi}\Big(|F_3|^2 + |F_4|^2\Big)\nonumber -\Re \Bigg[2\cos{\theta_\pi}F_1^* F_2 \nonumber \\ 
    &- \sin^2{\theta_\pi}\Big(F_1^* F_4 + F_2^* F_3 + \cos{\theta_\pi}F_3^*F_4\Big) \Bigg] \label{eq: H11 + H22 exc},
    \end{align}
    
    \noindent and
    \begin{align}
    \frac{1}{2}\sum_{\lambda_i, \lambda_f}H^{00} = &\frac{1}{2}\sin^2{\theta_\pi} \Big(|F_7|^2 + |F_8|^2 \nonumber \\ 
    &+ 2\cos{\theta_\pi}\Re \big(F_8^* F_7 \big)\Big), \label{eq: H00 exc}
    \end{align}

\noindent where the helicity of the final nucleon was summed over, and the helicity of the initial nucleon was averaged. The relations between the CGLN amplitudes and the multipoles can be used to obtain, after integrating over $\theta_\pi$, the expressions given in Eqs. (\ref{eq: H11 + H22 incl}) and (\ref{eq: H00 incl}).\\

\vspace{-0.5 cm}

\section{$K$- and $T$-matrix of the background contributions}\label{sec: App G}

\noindent The Born contributions of the background are computed from the HNV-Lagrangian \cite{Hernandez_HNVoffnucleon}, which contains nucleon and pion fields. For single-pion production, only the diagrams that are lowest order in the pion coupling are taken into account. In SPP, the final nucleon and pion can still interact with each other. Only considering strong processes and neglecting EM and weak processes in these final state interactions results in a set of coupled-channel equations \cite{DCC_coupled_channel_eq}

\begin{align}
    &T_{\pi \gamma}(k_\pi, q;W) = T^0_{\pi \gamma}(k_\pi, q; W)\nonumber \\
    + &\sum_\alpha \int p^2 dp \; T_{\pi \alpha}(k_\pi, p;W) G_\alpha(p; W)T^0_{\alpha \gamma}(p, q;W),
\end{align}

\vspace{-0.3cm}

\noindent where $\alpha$ denotes all intermediate hadronic states and the $T$-matrix elements are defined in Section \ref{sec: K matrix theory}. $T^0_{\pi\gamma}(p_1, p_2; W)$ represents the contribution from the Born terms. $G_\alpha(p; W)$ is the propagator for the intermediate state $\alpha$ with CMS momentum $p$. Below the two-pion production threshold, only the state containing one pion and one nucleon is energetically accessible. The coupled channel equation then becomes

\begin{align}
    &T_{\pi \gamma}(k_\pi, q;W) = T^0_{\pi \gamma}(k_\pi, q;W)\nonumber \\
    + &\int dp \; \frac{p^2 T_{\pi \pi}(k_\pi, p;W) T^0_{\pi \gamma}(p, q;W)}{W - E_\pi(p) - E_N(p) + i\epsilon},
\end{align}

\vspace{0.2 cm}

\noindent where $G_\alpha(p; W)$ was replaced by the pion-nucleon propagator. This propagator can be written as

\begin{align}
    &\frac{1}{W - E_\pi(p) - E_N(p) + i\epsilon} = \mathcal{P} \frac{1}{W - E_\pi(p) - E_N(p)} \nonumber \\
     &- i\pi \delta(W - E_\pi(p) - E_N(p)),\label{eq: principle value}
\end{align}

\vspace{0.2 cm}

\noindent with $\mathcal{P}$ the principal value representing the inelastic rescattering of the final nucleon and pion. For values of $W$ in the Delta region these inelastic processes can be neglected, resulting in

\begin{align}
    T_{\pi \gamma}(k_\pi, q;W) = &T^0_{\pi \gamma}(k_\pi, q;W)\nonumber \\
    -&i\pi \int dp \;  p^2 T_{\pi \pi}(k_\pi, p;W)T^0_{\pi \gamma}(p, q;W) \nonumber \\
    \times  &\delta(W - E_\pi(p) - E_N(p)).
\end{align}

\noindent The relation

\begin{equation}
    \int dx \; g(x)\delta(a - f(x)) = \frac{g(x)}{\frac{\partial f}{\partial x}} \Bigg|_{f(x) = a},
\end{equation}

\noindent can be used to write

\begin{align}
    &i\pi \int dp \; p^2 T_{\pi \pi}(k_\pi, p;W)T^0_{\pi \gamma}(p, q;W)\nonumber\\
    &\times \delta(W - E_\pi(p) - E_N(p)) \nonumber \\
    = &i\pi k_\pi^2 \frac{T_{\pi \pi}(k_\pi, k_\pi;W)T^0_{\pi \gamma}(k_\pi, q;W)}{-\frac{k_\pi}{E_\pi} - \frac{k_\pi}{E_N}}\\
    = &-i\pi \frac{k_\pi E_\pi E_N}{W}T_{\pi \pi}(k_\pi, k_\pi;W)T^0_{\pi \gamma}(k_\pi, q;W)\\
    = &-i \rho_\pi T_{\pi \pi}(k_\pi, k_\pi;W)T^0_{\pi \gamma}(k_\pi, q;W),
\end{align}

\noindent where the definition of the phase-space factor of Eq. (\ref{eq: phase-space}) is used. Note that the $\delta$-function constrains the CMS momentum of the intermediate pion–nucleon state to $p = k_\pi$. As a result, the pion–nucleon rescattering is elastic within this approximation, which follows from neglecting the principal-value contribution in Eq.~(\ref{eq: principle value}). The coupled-channel equation finally results in

\begin{equation}
T_{\pi\gamma} = T^0_{\pi\gamma} + i\rho_\pi T_{\pi\pi}T^0_{\pi\gamma},
\end{equation}

\noindent where the dependences on momenta and energy is omitted in the notation. Comparing this to the expression of the $T$-matrix in terms of the $K$-matrix

\begin{equation}
T_{\pi\gamma} = K_{\pi\gamma}\left(1 + i\rho_\pi T_{\pi\pi} \right),
\end{equation}

\noindent shows that identifying $K_{\pi\gamma}$ with $T^0_{\pi\gamma}$ corresponds to neglecting the principal value of the propagator. This last assumption is justified as in the Delta resonance region the scattering amplitude for pion-nucleon scattering remains elastic.

\section{$K$- and $T$-matrix of the Delta Resonance}\label{sec: App H}

 \noindent This Appendix provides a description of the Delta resonance within the framework of $K$-matrix theory. A resonance contribution to the scattering operator can be modeled as a pole in the $K$-matrix. In the case of a multipole involving a Delta contribution, the $K$-matrix elements are assumed to take the following form \cite{Workman_K_matrix_multiple_channels}

\vspace{-0.3 cm}

\begin{align}
    &K^\Delta_{\gamma \pi} = \frac{A_\Delta}{W^2 - M_\Delta^2} + B_\Delta, \label{eq: K delta 1} \\
    &K^\Delta_{\pi \pi} = \frac{C_\Delta}{W^2 - M_\Delta^2} + D_\Delta,\label{eq: K delta 2}
\end{align}

\noindent which includes a pole and background contribution. $A_\Delta,$ and $C_\Delta$ are related to the creation and decay vertices of the Delta while $B_\Delta$ and $D_\Delta$ contain the vertices of the background terms. Eqs.~(\ref{eq: K from unit}), (\ref{eq: T below 2pi}) and (\ref{eq: K below 2pi}) yield the following relations

\begin{align}
    -&iK_{\gamma \pi} T_{\gamma \pi} + T_{\pi \pi}(1-iK_{\pi \pi}) = K_{\pi \pi},\\
    &T_{\gamma \pi} = K_{\gamma \pi} (1 + i T_{\pi \pi}).
\end{align}

\noindent These equations can be combined to obtain the $T$-matrix element for SPP

\begin{align}
    T_{\gamma \pi} = \frac{K_{\gamma \pi}}{1 - iK_{\pi \pi} - iK_{\gamma \pi}^2}.
\end{align}

\noindent In the denominator, $K_{\gamma \pi}^2$ can be neglected because this matrix element is proportional to the square of the electroweak coupling constant. Inserting Eqs.~(\ref{eq: K delta 1}) and (\ref{eq: K delta 2}) results in

\begin{align}
    T^\Delta_{\gamma \pi} = &\left(\frac{A_\Delta}{W^2 - M_\Delta^2} + B_\Delta \right)\nonumber \\
    \times &\frac{(W^2 - M_\Delta^2)}{W^2 - M_\Delta^2 - i[C_\Delta + D_\Delta(W^2 - M_\Delta^2)]}. \label{eq: T matrix of delta}
\end{align}

\noindent This expression has a resonant part and a background part that goes to zero for $W = M_\Delta$. The complex part in the denominator is now related to the decay width of the Delta and it is clear that this width is determined by the dynamics of pion-nucleon scattering.\\

\noindent As a final remark, we mention that the separation in a resonant and non-resonant part is only unique at the pole. At any other energy, one can redefine $A_\Delta$ and $C_\Delta$ to absorb contributions from the non-resonant terms, so that the distinction between the two becomes ambiguous.

\bibliographystyle{apsrev4-1}
\bibliography{my_bibliography}

\end{document}